\documentclass[12pt, thmsb]{article}
\usepackage{amsfonts}

\usepackage{graphicx}
\usepackage{sw20lart}

\input{tcilatex}

\begin{document}

\title{A Hyperbolic System in a One-Dimensional Network}
\begin{center}
{\LARGE A Hyperbolic System in a One-Dimensional Network}

\bigskip

\textbf{Weihua Ruan}$^{\dagger }$\textbf{, M.E. Clark}$^{\ddagger }$\textbf{%
, Meide Zhao}$^{\ddagger }$\textbf{\ and Anthony Curcio}$^{\ddagger }$

\medskip

$^{\dagger }$\textbf{Department of Mathematics, Computer Science and
Statistics,}

\textbf{Purdue University Calumet}

\smallskip

\textbf{and}

$^{\ddagger }$\textbf{VasSol, Inc.}
\end{center}

\smallskip

\paragraph{Abstract.}

We study a coupled system of Navier-Stokes equation and the equation of
conservation of mass in a one-dimensional network. The system models the
blood circulation in arterial networks. A special feature of the system is
that the equations are coupled through boundary conditions at joints of the
network. We prove the existence and uniqueness of the solution to the
initial-boundary value problem, discuss the continuity of dependence of the
solution and its derivatives on initial, boundary and forcing functions and
their derivatives, develop a numerical scheme that generates discretized
solutions, and prove the convergence of the scheme.

\smallskip \bigskip

\section{\protect\smallskip Introduction\label{Int}}

\smallskip In this paper, we study a system of first-order quasilinear
hyperbolic partial differential equations defined on one-dimensional
networks. By network, we mean a finite collection of smooth curves with
finitely many intersections and endpoints. The mathematical system arises
from a long time study of fluid dynamical models that simulate blood flow in
arterial networks (cf. \cite{CK78,KC85,PYR86,RJS71,RJS74}). Recently, the
models have been used in technologies for medical diagnostics (\cite
{AB97,CC96,CMC97,CZL99}). In particular, a technology called CANVAS,
Computer-Assisted Non-invasive Vascular Analysis and Simulation, has been
developed to help stroke patients. CANVAS uses data from magnetic resonance
imaging to determine volumetric flow within vessels in the patient's brain
\cite{ZCA00}. The vessel flows were used to determine the boundary
conditions of the model \cite{CZL99}. It is based on a model formulated by
Clark and Kufahl \cite{CK78,KC85}. The technology has displayed its
capability in helping doctors predict outcomes of major medical procedures.
It is the extensive applications of these models that motivate their
mathematical study. Of particular importance are whether the mathematical
system is well-posed (solution exists, is unique, and is stable), and
whether the solutions generated by the computer algorithm really approximate
the true solutions.
\begin{figure}
\centering
\includegraphics{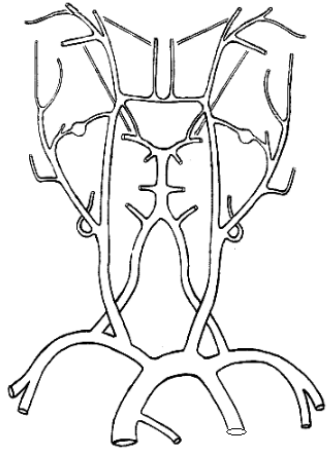}
\caption{A schematic diagram of an arterial network}
\end{figure}

In this paper, we study a generalization of a model given by \cite
{PYR86,RJS71,RJS74}, prove the existence and uniqueness of the
solution, prove the continuous dependence of the solution on the
initial, boundary, and forcing functions, and develop a numerical
scheme that approximates the solution.

To explain our system, let us first describe the original model of \cite
{PYR86,RJS71,RJS74}. Suppose an arterial network consists of $n$ vessels. We
parameterize each vessel with a spatial variable $x\in \left( 0,1\right) $.
In the vessel, the flow of blood is governed by conservation of mass and
Navier-Stokes momentum:
\begin{equation}
\begin{array}{l}
\displaystyle\frac{\partial Q_{i}}{\partial x}+\frac{\partial A_{i}}{%
\partial t}=0 \\[12pt]
\displaystyle\frac{\partial Q_{i}}{\partial t}+\frac{\partial }{\partial x}%
\left( \frac{Q_{i}^{2}}{A_{i}}\right) =-\frac{A_{i}}{\rho _{i}}\frac{%
\partial P_{i}}{\partial x}-\frac{8\pi \mu _{i}Q_{i}}{\rho _{i}A_{i}},
\end{array}
\quad x\in \left( 0,1\right) ,\ t>0,  \label{deaq}
\end{equation}
where $Q_{i}$ is the flow rate, $P_{i}$ is the pressure, $A_{i}$ is the
cross-sectional area of the vessel, and $\rho _{i}$, $\mu _{i}$ are positive
constants. The initial conditions are given by
\[
P_{i}\left( 0,x\right) =P_{i}^{I}\left( x\right) ,\quad Q_{i}\left(
0,x\right) =Q_{i}^{I}\left( x\right) ,\quad i=1,\ldots ,n.
\]
At each end of the vessel, depending on whether it is a source, an internal
junction, or a terminal, a boundary condition is imposed. At a source end,
either the pressure
\begin{equation}
P_{i}\left( 0,t\right) =P_{i}^{B}\left( t\right)  \label{bcsp}
\end{equation}
or the flow
\begin{equation}
Q_{i}\left( 0,t\right) =Q_{i}^{B}\left( t\right)  \label{bcsq}
\end{equation}
is specified. Various source ends may have different types of boundary
conditions. At an internal junction, suppose $j_{1},\ldots ,j_{\nu }$ are
the incoming vessels and $j_{\nu +1},\ldots ,j_{\mu }$ are the outgoing
vessels to the junction. We have mass and pressure continuities at junction
given by
\begin{equation}
\begin{array}{l}
\sum_{l=1}^{\nu }Q_{j_{l}}\left( 1,t\right) =\sum_{l^{\prime }=\nu +1}^{\mu
}Q_{j_{l^{\prime }}}\left( 0,t\right) , \\[12pt]
P_{j_{l}}\left( 1,t\right) =P_{j_{l^{\prime }}}\left( 0,t\right) ,\quad
1\leq l\leq \nu ,\ \nu +1\leq l^{\prime }\leq \mu .
\end{array}
\label{bcj}
\end{equation}
At a terminal end, we may specify either the pressure,

\begin{equation}
P_{i}\left( 1,t\right) =P_{i}^{B}\left( t\right) ,  \label{bctp}
\end{equation}
the flow,
\begin{equation}
Q_{i}\left( 1,t\right) =Q_{i}^{B}\left( t\right) ,  \label{bctq}
\end{equation}
or the impedance. In the last case, the boundary condition takes the form

\begin{equation}
\frac{\partial P_{i}}{\partial t}-\eta _{i}\frac{\partial Q_{i}}{\partial t}%
+\delta _{i}P_{i}-\varepsilon _{i}Q_{i}=W_{i}^{B}\left( t\right) ,\quad x=1,
\label{bctw}
\end{equation}
where $\eta _{i}$, $\delta _{i}$, and $\varepsilon _{i}$ are
positive constants and $W_{i}^{B}$ is a continuous function. This
equation arises from the windkessel model of peripheral bed, which
simulates the peripheral bed by a circuit that consists of a
resistance $R_{i}^{1}$ in series with the parallel combination of
a resistance $R_{i}^{2}$ and a capacitor $C_{i}$
\cite{KC85,PYR86,RJS74}. (See the diagram below.)
\begin{center}
\includegraphics{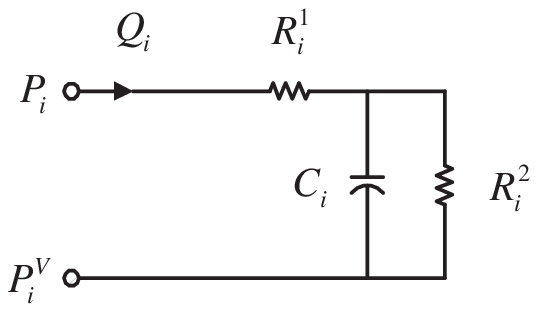}
\makeatletter\def\@captype{figure}\makeatother \caption{Electric
analog of the terminal boundary condition}
\end{center}

\noindent The resulting equation is
\[
\smallskip C_{i}\frac{\partial }{\partial t}\left( P_{i}-P_{i}^{V}\right)
-R_{i}^{1}C_{i}\frac{\partial Q_{i}}{\partial t}+\frac{P_{i}-P_{i}^{V}}{%
R_{i}^{2}}-\left( 1+\frac{R_{i}^{1}}{R_{i}^{2}}\right) Q_{i}=0,
\]
where $P_{i}^{V}$ is the venous pressure. It can be rewritten into (\ref
{bctw}). \noindent Again, boundary conditions for different terminals need
not be the same.

Finally, the cross-sectional area $A_{i}$ of the $i$-th vessel is a function
of $x$ and $P_{i}$. A particular example used in \cite{CK78,KC85} is
\[
A_{i}\left( x,P_{i}\right) =A_{i}^{0}\left( x\right) +\beta \ln \frac{P_{i}}{%
P_{i}^{0}}
\]
where $\beta $ is a positive constant and $A_{i}^{0}$ is a positive function
which represents the cross-sectional area at certain constant pressure $%
P_{i}^{0}$. This equation is used in \cite{CK78,KC85}.

\smallskip In this paper, we study a more general system which consists of
the equations
\begin{equation}
\begin{array}{ll}
\begin{array}{l}
\displaystyle\frac{\partial P_{i}}{\partial t}+a_{i}\frac{\partial Q_{i}}{%
\partial x}=f_{i}, \\[12pt]
\displaystyle\frac{\partial Q_{i}}{\partial t}+b_{i}\frac{\partial P_{i}}{%
\partial x}+2c_{i}\frac{\partial Q_{i}}{\partial x}=g_{i},
\end{array}
& \quad x\in \left( 0,1\right) ,\ t>0
\end{array}
\label{depq}
\end{equation}
and the initial and boundary conditions described above. For convenience, we
also use the vector form
\begin{equation}
\left( U_{i}\right) _{t}+B_{i}\left( U_{i}\right) _{x}=F_{i}  \label{devec}
\end{equation}
where $U_{i}=\left( P_{i},Q_{i}\right) $, $F_{i}=\left( f_{i},g_{i}\right) $
and
\[
B_{i}=\left(
\begin{array}{ll}
0 & a_{i} \\
b_{i} & 2c_{i}
\end{array}
\right) .
\]
Eq. (\ref{deaq}) is a special case of this system where
\[
a_{i}=\frac{1}{\left( A_{i}\right) _{P_{i}}},\quad b_{i}=\frac{A_{i}}{\rho
_{i}}-\frac{Q_{i}^{2}\left( A_{i}\right) _{P_{i}}}{A_{i}^{2}},\quad c_{i}=%
\frac{Q_{i}}{A_{i}},\ f_{i}=0,\ g_{i}=\frac{Q_{i}^{2}\left( A_{i}\right) _{x}%
}{A_{i}^{2}}-\frac{8\pi \mu _{i}Q_{i}}{\rho A_{i}}.
\]
We do not assume any particular form of these functions though, they are
general differentiable functions of $\left( x,t,P_{i},Q_{i}\right) $. A
basic assumption is $a_{i}>0$. Other assumptions will follow.

This problem is interesting not only in fluid mechanics but also in
mathematics. Navier-Stokes equations and conservation laws have been studied
for over a century. However, rarely have any studies been conducted for
systems defined in a network. Unlike the problem of fluid flow in a rigid
tube network, the distensibility of vessels greatly increases the complexity
of the problem. For example, as is well-known, a first-order quasilinear
system of hyperbolic equations on a finite one-dimensional spatial interval
needs not have a solution. Even if it has a solution for an interval of
time, the solution may not exist for all time. In a network, it is important
to know whether the coupling at junctions poses problems to solvability. The
effect of the windkessel boundary condition (\ref{bctw}) on the solvability
also needs to be examined.

This paper is divided into two parts. The first part consists of sections 2
and 3. It deals with the problem of solvability using a fixed point
approach. Substituting a pair of functions $\left( p_{i},q_{i}\right) $ for $%
\left( P_{i},Q_{i}\right) $ in the coefficients $a_{i}$, $b_{i}$, $c_{i}$
and forcing functions $f_{i}$, $g_{i}$, the system becomes linear. That is,
all the functions $a_{i}$, etc. are independent of unknowns. If the linear
system has a unique solution, then one can establish a mapping from $\left(
p_{i},q_{i}\right) $ to the linear problem solution $\left(
P_{i},Q_{i}\right) $. If one also shows that this mapping has a unique fixed
point, then the fixed point is necessarily the unique solution of the
quasilinear system. Hence, we shall first give a condition for the linear
system to have a unique solution, then examine under what conditions the
mapping has a unique fixed point. We investigate the first aspect of the
problem in Section 2 and the latter in Section 3. We also prove a result on
the continuity of dependence of solutions on the initial, boundary and
forcing functions for linear and quasilinear systems. Thus, we complete the
analysis of the well-posedness of the problem. In the second part, which
consists of Section 4 only, we give a numerical scheme that approximates the
solution, and prove its convergence. Our scheme is a set of
finite-difference equations based on the normal form of the differential
equations. Although these approaches are standard in the analysis of
quasilinear equations, the network feature of the system and the
peculiarities of the boundary conditions make the problem more complicated.
In the final section, we give a short discussion.

\section{The linear system}

\setcounter{equation}{0} \setcounter{theorem}{0}In this section, we analyze (%
\ref{depq}) as a linear system with $a_{i}$, $b_{i}$, $c_{i}$, $f_{i}$ and $%
g_{i}$ independent of $P_{i}$ and $Q_{i}$. We give conditions for the system
to have a unique global solution. The conditions are most naturally given in
terms of the eigenvalues of the matrix $B_{i}$, which have the form
\[
\lambda _{i}^{R}=c_{i}+u_{i},\quad \lambda _{i}^{L}=c_{i}-u_{i},
\]
where
\[
u_{i}=\sqrt{c_{i}^{2}+a_{i}b_{i}}.
\]
These eigenvalues are real if
\begin{equation}
c_{i}^{2}+a_{i}b_{i}>0,\ x\in \left( 0,1\right) ,\ t>0,\ i=1,\ldots ,n.
\label{cond2}
\end{equation}
In this case,
\begin{equation}
\lambda _{i}^{R}\left( x,t\right) >0,\text{ }\lambda _{i}^{L}\left(
x,t\right) <\lambda _{i}^{R}\left( x,t\right)  \label{linth1.9}
\end{equation}
and the system is hyperbolic. Under this condition, we show that the linear
system has a unique solution if
\[
\lambda _{i}^{L}\left( 0,t\right) <0,\ \lambda _{i}^{L}\left( 1,t\right)
<0,\ i=1,\ldots ,n.
\]
This is clearly equivalent to
\begin{equation}
a_{i}b_{i}>0,\ t\geq 0,\ i=1,\ldots ,n.  \label{cond3}
\end{equation}
at $x=0,1$ only. It needs not hold for $x\in \left( 0,1\right) $.

\begin{theorem}
\smallskip \label{linth1}Assume that the functions $a_{i}$, $b_{i}$, $c_{i}$%
, $f_{i}$, and $g_{i}$ are independent of $\left( P_{i},Q_{i}\right) $.
Suppose these functions and the initial and boundary functions $P_{i}^{I}$, $%
Q_{i}^{I}$, $P_{i}^{B}$, $Q_{i}^{B}$ and $W_{i}^{B}$ all have bounded
first-order derivatives. Suppose also that $a_{i}>0$ and that the conditions
(\ref{cond2}) and (\ref{cond3}) hold. Then, for any $T>0$ there is a unique
solution in a bounded subset of the space $C\left( \left[ 0,1\right] \times
\left[ 0,T\right] ,\Bbb{R}^{2n}\right) $ to the linear system (\ref{depq})
with the initial and boundary conditions given in Section \ref{Int}.
\end{theorem}

\paragraph{Proof.}

\smallskip We first show that the system has a unique solution for $%
0<t<\delta $ for some $\delta >0$. The proof is based on the method of
characteristics and a fixed point principle. For systems defined on only one
branch, this is a standard approach. In our case, special care is needed to
handle the junction condition (\ref{bcj}) and the windkessel boundary
condition (\ref{bctw}).

Consider the $i$-th branch. From any point $\left( \tau ,\xi \right) $ on
the left, right, and lower boundary of the rectangle $D=:\left[ 0,1\right]
\times \left[ 0,T\right] $, we construct the left-going and right-going
characteristic curves $x=x_{i}^{L}\left( t;\xi ,\tau \right) $ and $%
x=x_{i}^{R}\left( t;\xi ,\tau \right) $ by
\begin{eqnarray*}
\frac{dx_{i}^{L}}{dt} &=&\lambda _{i}^{L}\left( x_{i}^{L},t\right) ,\
x_{i}^{L}\left( \tau \right) =\xi , \\
\frac{dx_{i}^{R}}{dt} &=&\lambda _{i}^{R}\left( x_{i}^{R},t\right) ,\
x_{i}^{R}\left( \tau \right) =\xi ,
\end{eqnarray*}
respectively, where $\lambda _{i}^{L}$ and $\lambda _{i}^{R}$ are the two
eigenvalues of the matrix $B_{i}$. By the uniqueness of solutions of these
differential equations, a left-going (resp. right-going) characteristic
curve cannot intersect with another left-going (resp. right-going)
characteristic curve. Let $X_{i}^{L}$ and $X_{i}^{R}$ be the right-most
left-going and left-most right-going characteristic curves:
\[
x=x_{i}^{L}\left( t;1,0\right) \ \text{and }x=x_{i}^{R}\left( t;0,0\right)
\]
starting from the lower boundary of $D$, respectively. It can be shown from (%
\ref{linth1.9}) that the two curves can have at most one intersection. Let $%
t_{i}$ be the value of $t$ at the intersection. If the two curves do not
intersect in $D$, we simply define $t_{i}=T$. By condition (\ref{cond3}), $%
X_{i}^{L}$ cannot reach the right vertical line $x=1$ at any $t>0$, and by $%
\lambda _{i}^{R}>0$, $X_{i}^{R}$ cannot reach the vertical line $x=0$ at any
$t>0$. Thus, the rectangle $D_{i}=:\left[ 0,1\right] \times \left[
0,t_{i}\right] $ can be divided into three parts
\[
D_{i}=D_{i}^{L}\cup D_{i}^{C}\cup D_{i}^{R},
\]
where $D_{i}^{L}$ is between the vertical line $x=0$ and the characteristic
curve $X_{i}^{R}$, $D_{i}^{C}$ is between the two characteristic curves, and
$D_{i}^{R}$ is between $X_{i}^{L}$ and $x=1$.
\begin{center}
\includegraphics{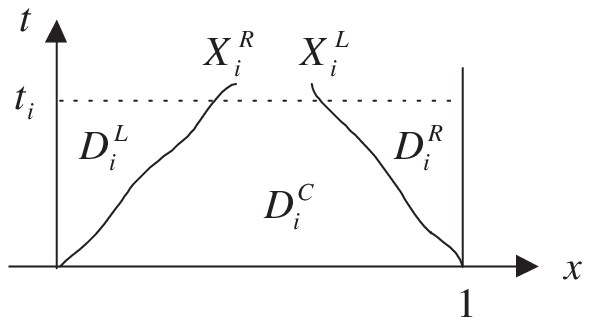}
\makeatletter\def\@captype{figure}\makeatother \caption{Three
parts of $D_{i}$}
\end{center}
\noindent We
show that there is a $\delta _{i}\leq t_{i}$ such that the solution $\left(
P_{i},Q_{i}\right) $ for the $i$-th branch exists in the restriction of $%
D_{i}$ to the strip $\left\{ 0\leq t\leq \delta _{i}\right\} $.

We first observe that the initial conditions alone determine the solution
completely in the central region $D_{i}^{C}$. This follows from the theory
of first-order linear hyperbolic systems and the fact that from any point $%
\left( x,t\right) \in D_{i}^{C}$, the two characteristic curves, followed
backwards, must land on the horizontal line $t=0$. (The latter is a
consequence of (\ref{linth1.9}).) To extend the solution to other parts of $%
D_{i}$, we make a change of unknowns and derive a set of integral equations.
Note that $l_{i}^{R}=:\left( -\lambda _{i}^{L},a_{i}\right) $ and $%
l_{i}^{L}=:\left( -\lambda _{i}^{R},a_{i}\right) $ are the left eigenvectors
of $B_{i}$ corresponding to $\lambda _{i}^{R}$ and $\lambda _{i}^{L}$,
respectively. Introduce new unknowns
\begin{equation}
r_{i}=l_{i}^{R}U_{i}\equiv -\lambda _{i}^{L}P_{i}+a_{i}Q_{i},\quad
s_{i}=l_{i}^{L}U_{i}\equiv -\lambda _{i}^{R}P_{i}+a_{i}Q_{i}.
\label{linth1.6}
\end{equation}
The system (\ref{depq}) can be written in terms of $r_{i}$ and $s_{i}$ by
multiplying the left eigenvectors to (\ref{devec}) and substituting in
\begin{equation}
P_{i}=\frac{1}{2u_{i}}\left( r_{i}-s_{i}\right) ,\quad Q_{i}=\frac{1}{%
2u_{i}a_{i}}\left( \lambda _{i}^{R}r_{i}-\lambda _{i}^{L}s_{i}\right) .
\label{linth1.1}
\end{equation}
This results in the equations
\begin{equation}
\partial _{i}^{R}r_{i}=F_{i}^{R}\left( x,t,r_{i},s_{i}\right) ,\ \partial
_{i}^{L}s_{i}=F_{i}^{L}\left( x,t,r_{i},s_{i}\right) ,  \label{linth1.2}
\end{equation}
where
\begin{equation}
\partial _{i}^{R}=\frac{\partial }{\partial t}+\lambda _{i}^{R}\frac{%
\partial }{\partial x},\quad \partial _{i}^{L}=\frac{\partial }{\partial t}%
+\lambda _{i}^{L}\frac{\partial }{\partial x},  \label{linth1.5}
\end{equation}
and
\begin{equation}
F_{i}^{R}\left( x,t,r_{i},s_{i}\right) =l_{i}^{R}F_{i}+\left( \partial
_{i}^{R}l_{i}^{R}\right) U_{i},\quad F_{i}^{L}\left( x,t,r_{i},s_{i}\right)
=l_{i}^{L}F_{i}+\left( \partial _{i}^{L}l_{i}^{L}\right) U_{i}.
\label{linth1.3}
\end{equation}
(A differential operator acting on a vector means that it acts on each
component of the vector.) Let $\left( x,t\right) \in D_{i}$. We integrate
the first equation of (\ref{linth1.2}) along the right-going characteristic
curve $x^{R}\left( t;\xi ,\tau \right) $ which passes through $\left(
x,t\right) $ and reaches the left or lower boundary of $D_{i}$ at $\left(
\xi ,\tau \right) $. It can be shown that for $\left( x,t\right) \in
D_{i}^{C}\cup D_{i}^{R}$, $\tau =0$, and for $\left( x,t\right) \in
D_{i}^{L} $, $\xi =0$. In the former case, we obtain
\begin{equation}
r_{i}\left( x,t\right) =r_{i}^{I}\left( \xi \right)
+\int_{0}^{t}F_{i}^{R}\left( x_{i}^{R}\left( t^{\prime };\xi ,0\right)
,t^{\prime },r_{i},s_{i}\right) dt^{\prime }  \label{inteq1}
\end{equation}
In the latter case, we have
\begin{equation}
r_{i}\left( x,t\right) =r_{i}\left( 0,\tau \right) +\int_{\tau
}^{t}F_{i}^{R}\left( x_{i}^{R}\left( t^{\prime };0,\tau \right) ,t^{\prime
},r_{i},s_{i}\right) dt^{\prime }.  \label{inteq2}
\end{equation}
Similarly, by integrating the second equation of (\ref{linth1.2}) along the
left-going characteristic curve $x_{i}^{L}\left( t;\xi ,\tau \right) $ that
passes through both $\left( x,t\right) $ and $\left( \xi ,\tau \right) $
(which is on either the right or lower boundary of $D_{i}$), the equations
are
\begin{equation}
s_{i}\left( x,t\right) =s_{i}^{I}\left( \xi \right)
+\int_{0}^{t}F_{i}^{L}\left( x_{i}^{L}\left( t^{\prime };\xi ,0\right)
,t^{\prime },r_{i},s_{i}\right) dt^{\prime }  \label{inteq3}
\end{equation}
if $\left( x,t\right) \in D_{i}^{L}\cup D_{i}^{C}$ and
\begin{equation}
s_{i}\left( x,t\right) =s_{i}\left( 1,\tau \right) +\int_{\tau
}^{t}F_{i}^{L}\left( x_{i}^{L}\left( t^{\prime };1,\tau \right) ,t^{\prime
},r_{i},s_{i}\right) dt^{\prime }  \label{inteq4}
\end{equation}
if $\left( x,t\right) \in D_{i}^{R}$. These are the integral equations we
need.

For any $\delta _{i}\leq t_{i}$ we use $D_{i,\delta _{i}}^{L}$, $D_{i,\delta
_{i}}^{C}$ and $D_{i,\delta _{i}}^{R}$ to denote the restrictions of $%
D_{i}^{L}$, $D_{i}^{C}$ and $D_{i}^{R}$ to the strip $\left\{ 0\leq t\leq
\delta _{i}\right\} $, respectively. We first extend the solution to a left
region $D_{i,\delta _{i}}^{L}$ where $\delta _{i}$ is to be determined. For
this, we need the boundary condition on the left end of the branch. The left
end is either a source or a junction. For a source with the boundary
condition (\ref{bcsp}), we define $\hat{s}_{i}=s_{i}/\varepsilon $ where $%
\varepsilon <1$ is any constant. Using the first equation of (\ref{linth1.1}%
) in the integral equations (\ref{inteq2}) and (\ref{inteq3}),
\begin{equation}
\left(
\begin{array}{l}
r_{i}\left( x,t\right) \\
\hat{s}_{i}\left( x,t\right)
\end{array}
\right) =\left(
\begin{array}{l}
\displaystyle2u_{i}\left( 0,\tau \right) P_{i}^{B}\left( \tau \right)
+\varepsilon \hat{s}_{i}\left( 0,\tau \right) +\int_{\tau
}^{t}F_{i}^{R}\left( x_{i}^{R}\left( t^{\prime };0,\tau \right) ,t^{\prime
},r_{i},\varepsilon \hat{s}_{i}\right) dt^{\prime } \\
\displaystyle\frac{1}{\varepsilon }s_{i}^{I}\left( \xi \right) +\frac{1}{%
\varepsilon }\int_{0}^{t}F_{i}^{L}\left( x_{i}^{L}\left( t^{\prime };\xi
,0\right) ,t^{\prime },r_{i},\varepsilon \hat{s}_{i}\right) dt^{\prime }
\end{array}
\right) .  \label{inteq5}
\end{equation}
This is a fixed point equation for $\left( r_{i},\hat{s}_{i}\right) $ if we
define the right hand side as a mapping of an operator $K$ on $\left( r_{i},%
\hat{s}_{i}\right) $ in a bounded subset of $C\left( D_{i,\delta
_{i}}^{L}\cup D_{i,\delta _{i}}^{C},\Bbb{R}^{2}\right) $. In a standard
approach, it can be shown that $K$ is a contraction mapping if $\delta _{i}$
is sufficiently small. Hence, the fixed point exists and is unique.
Therefore, the solution $\left( r_{i},s_{i}\right) $ can be uniquely
extended to $D_{i,\delta _{i}}^{L}\cup D_{i,\delta _{i}}^{C}$.

For a source with the boundary condition (\ref{bcsq}), we define $\hat{s}%
_{i}=s_{i}/\varepsilon $, where $\varepsilon >0$ and is so small such that
\[
\varepsilon \left| \frac{\lambda _{i}^{L}\left( 0,\tau \right) }{\lambda
_{i}^{R}\left( 0,\tau \right) }\right| <1,\quad \tau \in \left(
0,t_{i}\right) .
\]
The fixed point equation is then
\begin{equation}
\left(
\begin{array}{l}
r_{i}\left( x,t\right) \\
\hat{s}_{i}\left( x,t\right)
\end{array}
\right) =\left(
\begin{array}{l}
\displaystyle\frac{2a_{i}u_{i}\left( 0,\tau \right) }{\lambda _{i}^{R}\left(
0,\tau \right) }Q_{i}^{B}\left( \tau \right) +\frac{\lambda _{i}^{L}\left(
0,\tau \right) }{\lambda _{i}^{R}\left( 0,\tau \right) }\varepsilon \hat{s}%
_{i}\left( 0,\tau \right) +\int_{\tau }^{t}F_{i}^{R}\left( x_{i}^{R}\left(
t^{\prime };0,\tau \right) ,t^{\prime },r_{i},\varepsilon \hat{s}_{i}\right)
dt^{\prime } \\
\displaystyle\frac{1}{\varepsilon }s_{i}^{I}\left( \xi \right) +\frac{1}{%
\varepsilon }\int_{0}^{t}F_{i}^{L}\left( x_{i}^{L}\left( t^{\prime };\xi
,0\right) ,t^{\prime },r_{i},\varepsilon \hat{s}_{i}\right) dt^{\prime }
\end{array}
\right) .  \label{inteq5'}
\end{equation}
By a similar argument, the solution can again be uniquely extended.

If the left end of the branch is a junction, we shall extend the solution on
all the branches that are connected to the same junction simultaneously.
Thus, also extend the solution to $D_{i,\delta _{i}}^{R}$ on the branches
incoming to the junction. Let $j_{1},\ldots ,j_{\nu }$ be the incoming and $%
j_{\nu +1},\ldots ,j_{\mu }$ the outgoing branches to the junction.
Equations (\ref{bcj}) and (\ref{linth1.1}) give rise to a $2\mu \times \mu $
homogenous system of linear equations for $r_{i}\left( 1,\tau \right) $, $%
s_{i}\left( 1,\tau \right) $, $i=j_{1},\ldots ,j_{\nu }$ and $r_{i}\left(
0,\tau \right) $, $s_{i}\left( 0,\tau \right) $, $i=j_{\nu +1},\ldots
,j_{\mu }$:
\[
\begin{array}{l}
\frac{1}{u_{1}\left( 1,\tau \right) }\left( r_{1}\left( 1,\tau \right)
-s_{1}\left( 1,\tau \right) \right) -\frac{1}{u_{i}\left( 1,\tau \right) }%
\left( r_{i}\left( 1,\tau \right) -s_{i}\left( 1,\tau \right) \right)
=0,\quad i=j_{2},\ldots ,j_{\nu }, \\[12pt]
\frac{1}{u_{1}\left( 1,\tau \right) }\left( r_{1}\left( 1,\tau \right)
-s_{1}\left( 1,\tau \right) \right) -\frac{1}{u_{i}\left( 0,\tau \right) }%
\left( r_{i}\left( 0,\tau \right) -s_{i}\left( 0,\tau \right) \right)
=0,\quad i=j_{\nu +1},,\ldots ,j_{\mu }, \\[12pt]
\sum_{l=1}^{\nu }\frac{1}{u_{j_{l}}a_{j_{l}}}\left( \lambda
_{j_{l}}^{R}r_{j_{l}}-\lambda _{j_{l}}^{L}s_{j_{l}}\right) \left( 1,\tau
\right) -\sum_{l^{\prime }=\nu +1}^{\mu }\frac{1}{u_{j_{l^{\prime
}}}a_{j_{l^{\prime }}}}\left( \lambda _{j_{l^{\prime }}}^{R}r_{j_{l^{\prime
}}}-\lambda _{j_{l^{\prime }}}^{L}s_{j_{l^{\prime }}}\right) \left( 0,\tau
\right) =0.
\end{array}
\]
This system can be solved for $s_{j_{1}}\left( 1,\tau \right) ,\ldots
,s_{j_{\nu }}\left( 1,\tau \right) $, $r_{j_{\nu +1}}\left( 0,\tau \right)
,\ldots ,r_{j_{\mu }}\left( 0,\tau \right) $ because the coefficient matrix
\[
\left(
\begin{array}{cccc}
-\frac{1}{u_{j_{1}}\left( 1,\tau \right) } & \frac{1}{u_{j_{2}}\left( 1,\tau
\right) } & \cdots & 0 \\
\vdots & \vdots & \ddots & \vdots \\
-\frac{1}{u_{j_{1}}\left( 1,\tau \right) } & 0 & \cdots & -\frac{1}{%
u_{j_{\mu }}\left( 0,\tau \right) } \\
-\frac{\lambda _{j_{1}}^{L}\left( 1,\tau \right) }{u_{j_{1}}a_{j_{1}}\left(
1,\tau \right) } & -\frac{\lambda _{j_{2}}^{L}\left( 1,\tau \right) }{%
u_{j_{2}}a_{j_{2}}\left( 1,\tau \right) } & \cdots & -\frac{\lambda _{j_{\mu
}}^{R}\left( 0,\tau \right) }{u_{j_{\mu }}a_{j_{\mu }}\left( 0,\tau \right) }
\end{array}
\right)
\]
has the determinant
\[
\frac{\left( -1\right) ^{\nu +1}}{\prod_{l=1}^{\nu }u_{j_{l}}\left( 1,\tau
\right) \prod_{l^{\prime }=\nu +1}^{\mu }u_{j_{l^{\prime }}}\left( 0,\tau
\right) }\left( -\sum_{l=1}^{\nu }\frac{\lambda _{j_{l}}^{L}\left( 1,\tau
\right) }{a_{j_{l}}\left( 1,\tau \right) }+\sum_{l^{\prime }=\nu +1}^{\mu }%
\frac{\lambda _{j_{l^{\prime }}}^{R}\left( 0,\tau \right) }{a_{j_{l^{\prime
}}}\left( 0,\tau \right) }\right) .
\]
Since $\lambda _{i}^{L}<0<\lambda _{i}^{R}$ at the junction, the determinant
is not zero. Hence, we can express $s_{j_{1}}\left( 1,\tau \right) ,\ldots
,s_{j_{\nu }}\left( 1,\tau \right) $, $r_{j_{\nu +1}}\left( 0,\tau \right)
,\ldots ,r_{j_{\mu }}\left( 0,\tau \right) $ in terms of other unknowns as
\begin{eqnarray*}
s_{i}\left( 1,\tau \right) &=&\sum_{l=1}^{\nu }m_{j_{l}}^{i}\left( \tau
\right) r_{j_{l}}\left( 1,\tau \right) +\sum_{l^{\prime }=\nu +1}^{\mu
}m_{j_{l^{\prime }}}^{i}\left( \tau \right) s_{j_{l^{\prime }}}\left( 0,\tau
\right) ,\quad i=j_{1},\ldots ,j_{\nu }, \\
r_{i}\left( 0,\tau \right) &=&\sum_{l=1}^{\nu }n_{j_{l}}^{i}\left( \tau
\right) r_{j_{l}}\left( 1,\tau \right) +\sum_{l^{\prime }=\nu +1}^{\mu
}n_{j_{l^{\prime }}}^{i}\left( \tau \right) s_{j_{l^{\prime }}}\left( 0,\tau
\right) ,\quad i=j_{\nu +1},\ldots ,j_{\mu }
\end{eqnarray*}
for some functions $m_{j}^{i}$, $n_{j}^{i}$. Choose an $\varepsilon >0$ such
that
\[
\varepsilon \max \left\{ \sum_{l=1}^{\mu }\left| m_{j_{l}}^{i}\left( \tau
\right) \right| ,\sum_{l=1}^{\mu }\left| n_{j_{l}}^{i}\left( \tau \right)
\right| \right\} <1,\quad i=j_{1},\ldots ,j_{\mu },\ \tau \in \left[
0,t_{i}\right]
\]
and introduce
\[
\hat{r}_{j_{l}}=\frac{r_{j_{l}}}{\varepsilon },\quad \hat{s}_{j_{l^{\prime
}}}=\frac{s_{j_{l^{\prime }}}}{\varepsilon },\quad l=1,\ldots ,\nu ,\
l^{\prime }=\nu +1,\ldots ,\mu .
\]
Then, from (\ref{inteq1})--(\ref{inteq4}), the integral equations for the $%
2\mu $ unknowns $\hat{r}_{j_{l}}$, $s_{j_{l}}$, $r_{j_{l^{\prime }}}$, $\hat{%
s}_{j_{l^{\prime }}}$, $l=1,\ldots ,\nu $, $l^{\prime }=\nu +1,\ldots ,\mu $
constitute a fixed point equation, $w=Kw$, where
\begin{equation}
w=\left( \hat{r}_{j_{1}},\ldots ,\hat{r}_{j_{\nu }},s_{j_{1}},\ldots
,s_{j_{\nu }},r_{j_{\nu +1}},\ldots ,r_{j_{\mu }},\hat{s}_{j_{\nu
+1}},\ldots ,\hat{s}_{j_{\mu }}\right)  \label{linth1.7}
\end{equation}
and
\begin{equation}
\begin{array}{ll}
Kw= & \left( \frac{1}{\varepsilon }r_{j_{1}}^{I}\left( \xi _{j_{1}}\right) +%
\frac{1}{\varepsilon }\int_{0}^{t}F_{j_{1}}^{R}\left(
x_{j_{1}}^{R},t^{\prime },\varepsilon \hat{r}_{j_{1}},s_{j_{1}}\right)
dt^{\prime },\ldots ,\right. \\
& \varepsilon \left( \sum_{k=1}^{\nu }m_{j_{k}}^{1}\hat{r}_{j_{k}}\left(
1,\tau \right) +\sum_{k^{\prime }=\nu +1}^{\mu }m_{j_{k^{\prime }}}^{1}\hat{s%
}_{j_{k^{\prime }}}\left( 0,\tau \right) \right) +\int_{\tau
}^{t}F_{j_{1}}^{L}\left( x_{j_{1}}^{L},t^{\prime },\varepsilon \hat{r}%
_{j_{1}},s_{j_{1}}\right) dt^{\prime },\ldots , \\
& \varepsilon \left( \sum_{k=1}^{\nu }n_{j_{k}}^{1}\hat{r}_{j_{k}}\left(
1,\tau \right) +\sum_{k^{\prime }=\nu +1}^{\mu }n_{j_{k^{\prime }}}^{1}\hat{s%
}_{j_{k^{\prime }}}\left( 1,\tau \right) \right) +\int_{\tau }^{t}F_{j_{\nu
+1}}^{R}\left( x_{j_{\nu +1}}^{R},t^{\prime },r_{j_{\nu +1}},\varepsilon
\hat{s}_{j_{\nu +1}}\right) dt^{\prime },\ldots , \\
& \left. \frac{1}{\varepsilon }s_{j_{\nu +1}}^{I}\left( \xi _{j_{\nu
+1}}\right) +\frac{1}{\varepsilon }\int_{0}^{t}F_{j_{\nu +1}}^{L}\left(
x_{j_{\nu +1}}^{L},t^{\prime },r_{j_{\nu +1}},\varepsilon \hat{s}_{j_{\nu
+1}}\right) dt^{\prime },\ldots \right) .
\end{array}
\label{linth1.8}
\end{equation}
It can be shown by a standard argument that $K$ is a contraction mapping in
the space
\[
X_{j}=:\prod_{l=1}^{\nu }C\left( D_{j_{l},\delta _{j}}^{C}\cup
D_{j_{l},\delta _{j}}^{R},\Bbb{R}^{2}\right) \times \prod_{l=\nu +1}^{\mu
}C\left( D_{j_{l},\delta _{j}}^{L}\cup D_{j_{l},\delta _{j}}^{L},\Bbb{R}%
^{2}\right)
\]
if $\delta _{j}$ is sufficiently small. Hence, it has a unique fixed point
in $X_{j}$. This extends the solution $\left( r_{i},s_{i}\right) $ for the
neighboring branches of the junction.

We now extend the solution $\left( r_{i},s_{i}\right) $ to a right region $%
D_{i,\delta _{i}}^{R}$. This has been done if the right end is a junction.
Thus, only terminal ends need to be discussed. For the boundary condition of
either (\ref{bctp}) or (\ref{bctq}) type, the argument is similar to the
above discussion about source ends. We only sketch the steps in these two
cases. The boundary condition of (\ref{bctw}) type, however, requires more
effort.

If condition (\ref{bctp}) is assumed, then, by (\ref{linth1.1}),
\[
s_{i}\left( 1,t\right) =r_{i}\left( 1,t\right) -2u_{i}P_{i}^{B}\left(
t\right) .
\]
Let $\hat{r}_{i}=r_{i}/\varepsilon $ with $0<\varepsilon <1$. Then, the
fixed point equation for $\left( \hat{r}_{i},s_{i}\right) $ has the form
\begin{equation}
\left(
\begin{array}{r}
\hat{r}_{i}\left( x,t\right) \\
s_{i}\left( x,t\right)
\end{array}
\right) =\left(
\begin{array}{l}
\displaystyle\frac{1}{\varepsilon }r_{i}^{I}\left( \xi \right) +\frac{1}{%
\varepsilon }\int_{0}^{t}F_{i}^{R}\left( t^{\prime },x_{i}^{R}\left(
t^{\prime };\xi ,0\right) ,\varepsilon \hat{r}_{i},s_{i}\right) dt^{\prime }
\\
\displaystyle \varepsilon \hat{r}_{i}\left( 1,\tau \right) -2u_{i}\left(
1,\tau \right) P_{i}^{B}\left( \tau \right) +\int_{\tau }^{t}F_{i}^{L}\left(
t^{\prime },x_{i}^{L}\left( t^{\prime };1,\tau \right) ,\varepsilon \hat{r}%
_{i},s_{i}\right) dt^{\prime }
\end{array}
\right) .  \label{inteq8}
\end{equation}
As before, the mapping defined by the right hand side is contractive if $%
\delta _{i}$ is small enough. Hence, the solution is uniquely extended into $%
D_{i,\delta _{i}}^{R}$. If condition (\ref{bctq}) is assumed, we find again
from (\ref{linth1.1}) that
\[
\lambda _{i}^{L}s_{i}\left( 1,t\right) =\lambda _{i}^{R}r_{i}\left(
1,t\right) -2u_{i}\left( 1,t\right) a_{i}\left( 1,t\right) Q_{i}^{B}\left(
t\right) .
\]
Since $\lambda _{i}^{L}\left( 1,t\right) <0$, the equation can be uniquely
solved for $s_{i}$. Choose $\varepsilon >0$ sufficiently small such that
\[
\varepsilon \left| \frac{\lambda _{i}^{R}\left( 1,\tau \right) }{\lambda
_{i}^{L}\left( 1,\tau \right) }\right| <1\quad \text{for }\tau \in \left[
0,t_{i}\right]
\]
and let $\hat{r}_{i}=r_{i}/\varepsilon $. The fixed point equation for $%
\left( \hat{r}_{i},s_{i}\right) $ has the form
\begin{equation}
\left(
\begin{array}{r}
\hat{r}_{i}\left( x,t\right) \\
s_{i}\left( x,t\right)
\end{array}
\right) =\left(
\begin{array}{l}
\displaystyle\frac{1}{\varepsilon }r_{i}^{I}\left( \xi \right) +\frac{1}{%
\varepsilon }\int_{0}^{t}F_{i}^{R}\left( x_{i}^{R}\left( t^{\prime };\xi
,0\right) ,t^{\prime },\varepsilon \hat{r}_{i},s_{i}\right) dt^{\prime } \\
\displaystyle \frac{\lambda _{i}^{R}\left( 1,\tau \right) }{\lambda
_{i}^{L}\left( 1,\tau \right) }\varepsilon \hat{r}_{i}\left( 1,\tau \right) -%
\frac{2a_{i}u_{i}\left( 1,\tau \right) }{\lambda _{i}^{L}\left( 1,\tau
\right) }Q_{i}^{B}\left( \tau \right) +\int_{\tau }^{t}F_{i}^{L}\left(
x_{i}^{L}\left( t^{\prime };1,\tau \right) ,t^{\prime },\varepsilon \hat{r}%
_{i},s_{i}\right) dt^{\prime }
\end{array}
\right) .  \label{inteq8'}
\end{equation}
Again, the mapping is contractive in a bounded subset of $C\left(
D_{i,\delta _{i}}^{C}\cup D_{i,\delta _{i}}^{R},\Bbb{R}^{2}\right) $ if $%
\delta _{i}$ is sufficiently small. The solution is thus, uniquely extended
to $D_{i,\delta _{i}}^{R}$.

In the case where the boundary condition (\ref{bctw}) is assumed, we
integrate it with respect to $t$ to obtain
\[
\left( P_{i}-\eta _{i}Q_{i}\right) \left( 1,t\right) =\left( P_{i}^{I}-\eta
_{i}Q_{i}^{I}\right) \left( 1\right) +\int_{0}^{t}\left( W_{i}^{B}\left(
t^{\prime }\right) -\delta _{i}P_{i}\left( 1,t^{\prime }\right) +\varepsilon
_{i}Q_{i}\left( 1,t^{\prime }\right) \right) dt^{\prime }.
\]
Substituting (\ref{linth1.1}) into this equation, we can write
\[
m_{i}\left( t\right) r_{i}\left( 1,t\right) -n_{i}\left( t\right)
s_{i}\left( 1,t\right) =m_{i}\left( 0\right) r_{i}^{I}\left( 1\right)
-n_{i}\left( 0\right) s_{i}^{I}\left( 1\right) +\int_{0}^{t}H_{i}\left(
t^{\prime },r_{i}\left( 1,t^{\prime }\right) ,s_{i}\left( 1,t^{\prime
}\right) \right) dt^{\prime }
\]
where
\[
m_{i}\left( t\right) =\frac{a_{i}\left( 1,t\right) -\eta _{i}\lambda
_{i}^{R}\left( 1,t\right) }{2a_{i}u_{i}\left( 1,t\right) },\quad n_{i}\left(
t\right) =\frac{-a_{i}\left( 1,t\right) +\eta _{i}\lambda _{i}^{L}\left(
1,t\right) }{2a_{i}u_{i}\left( 1,t\right) }
\]
and
\[
H_{i}\left( t,r,s\right) =W_{i}^{B}\left( t\right) +\frac{\varepsilon
_{i}\lambda _{i}^{R}\left( 1,t\right) -\delta _{i}a_{i}\left( 1,t\right) }{%
2a_{i}u_{i}\left( 1,t\right) }r-\frac{\varepsilon _{i}\lambda _{i}^{L}\left(
1,t\right) -\delta _{i}a_{i}\left( 1,t\right) }{2a_{i}u_{i}\left( 1,t\right)
}s.
\]
Since $a_{i}>0$, $u_{i}>0$, $\eta _{i}>0$ and $\lambda _{i}^{L}\left(
1,t\right) <0$, it follows that $n_{i}\left( t\right) <0$. Hence, there
exists $\varepsilon >0$ such that
\[
\varepsilon \left| \frac{m_{i}\left( \tau \right) }{n_{i}\left( \tau \right)
}\right| <1\quad \text{for }\tau \in \left[ 0,t_{i}\right] .
\]
Let $\hat{r}_{i}=r_{i}/\varepsilon $. The integral equations for $\hat{r}%
_{i} $ and $s_{i}$ then have the form
\begin{equation}
\begin{array}{l}
\displaystyle\hat{r}_{i}\left( x,t\right) =\frac{1}{\varepsilon }%
r_{i}^{I}\left( \xi \right) +\frac{1}{\varepsilon }\int_{0}^{t}F_{i}^{R}%
\left( x_{i}^{R}\left( t^{\prime };\xi ,0\right) ,t^{\prime },\varepsilon
\hat{r}_{i},s_{i}\right) dt^{\prime }, \\
\displaystyle s_{i}\left( x,t\right) =\varepsilon \frac{m_{i}\left( \tau
\right) }{n_{i}\left( \tau \right) }\hat{r}_{i}\left( 1,\tau \right) -\frac{1%
}{n_{i}\left( \tau \right) }\left( M_{i}+\int_{0}^{t}H_{i}\left( t^{\prime
},\varepsilon \hat{r}_{i}\left( 1,t^{\prime }\right) ,s_{i}\left(
1,t^{\prime }\right) \right) dt^{\prime }\right) \\
\hspace{2in}\displaystyle+\int_{\tau }^{t}F_{i}^{L}\left( x_{i}^{L}\left(
t^{\prime };1,\tau \right) ,t^{\prime },\varepsilon \hat{r}_{i},s_{i}\right)
dt^{\prime },
\end{array}
\label{inteq8''}
\end{equation}
where $M_{i}=m_{i}\left( 0\right) r_{i}^{I}\left( 1\right) -n_{i}\left(
0\right) s_{i}^{I}\left( 1\right) $ is a constant. The extension of the
solution to $D_{i,\delta _{i}}^{R}$ is thus, guaranteed.

Finally, if we let $\delta $ be the minimum of all $\delta _{i}$ occurring
above, we see that $\delta >0$ and the solution exists and is unique in $%
\left( x,t\right) \in D_{\delta }=:\left[ 0,1\right] \times \left[ 0,\delta
\right] $. Observe that $\delta $ depends only on the bounds of the system
functions $a_{i}$, etc., the initial and boundary functions $P_{i}^{I}$,
etc., and their first-order derivatives in $D=\left[ 0,1\right] \times
\left[ 0,T\right] $. Hence, it is independent of $t$, and we can extend the
solution successively in the time intervals $\left[ 0,\delta \right] $, $%
\left[ \delta ,2\delta \right] $, etc. In this way, the solution is obtained
in $D$ in finitely many steps. \endproof

\smallskip It can be seen from the above proof that the linear system needs
not have a solution if condition (\ref{cond3}) fails at any end point of a
branch. In the quasilinear case, since $a_{i}$ and $b_{i}$ depend on the
unknowns $P_{i}$ and $Q_{i}$, this condition may fail at a future moment.
Therefore the solution does not generally exist for all time.

We next derive an estimate of the deviation of solution in term of the
deviations of the initial, boundary and forcing functions. This estimate is
needed in the next section. For any vector function $v=\left( v_{1},\ldots
,v_{k}\right) $ defined in $C\left( X;\Bbb{R}^{k}\right) $, we use $\left|
v\right| _{X}$ to denote the norm $\max_{i}\left\{ \left| v_{i}\right|
_{C\left( X\right) }\right\} $, where $X$ represents a closed subset of
either $\Bbb{R}$ or $\Bbb{R}^{2}$.

\begin{lemma}
\label{conth1}Let $U=\left( P,Q\right) $ and $\tilde{U}=\left( \tilde{P},%
\tilde{Q}\right) $ be two solutions of the linear problem (\ref{devec}) with
different initial, boundary, and forcing functions. Suppose the conditions
of Theorem \ref{linth1} holds for both solutions. Then, there exists a
constant $M>0$, independent of initial, boundary and forcing functions, such
that
\begin{equation}
\begin{array}{r}
\left| U-\tilde{U}\right| _{C\left( D_{\delta }\right) }\leq M\left( \left|
P^{I}-\tilde{P}^{I}\right| _{C\left[ 0,1\right] }+\left| Q^{I}-\tilde{Q}%
^{I}\right| _{C\left[ 0,1\right] }+\left| P^{B}-\tilde{P}^{B}\right|
_{C\left[ 0,\delta \right] }+\left| Q^{B}-\tilde{Q}^{B}\right| _{C\left[
0,\delta \right] }\right. \\
+\left. \delta \left| f-\tilde{f}\right| _{C\left( D_{\delta }\right)
}+\delta \left| g-\tilde{g}\right| _{C\left( D_{\delta }\right) }+\delta
\left| W-\tilde{W}\right| _{C\left[ 0,\delta \right] }\right) .
\end{array}
\label{conth1.1}
\end{equation}
\end{lemma}

\paragraph{\noindent Proof.}

We need only prove (\ref{conth1.1}) for a $\delta \leq \min_{i}\left\{
\delta _{i}\right\} $, where $\delta _{i}$ represents the constants
occurring in the proof of Theorem \ref{linth1}. This is because for larger $%
\delta $, we can divide the interval $\left[ 0,\delta \right] $ into
subintervals, each has a length less than $\min_{i}\left\{ \delta
_{i}\right\} $, and apply (\ref{conth1.1}) in each subinterval. We can then
take the maximum on each side of the inequalities to derive the inequality
of in $\left[ 0,\delta \right] $. In the sequel, $D_{\delta }^{C}$, $%
D_{\delta }^{L}$ and $D_{\delta }^{R}$ are the restrictions of $D_{i}^{C}$, $%
D_{i}^{L}$ and $D_{i}^{R}$ to the strip $\left\{ 0\leq t\leq \delta \right\}
$, respectively.

By linearity, $U-\tilde{U}$ is the solution of the system with the initial,
boundary and forcing functions $P_{i}^{I}-\tilde{P}_{i}^{I}$, $Q_{i}^{I}-%
\tilde{Q}_{i}^{I}$, $P_{i}^{B}-\tilde{P}_{i}^{B}$, $Q_{i}^{B}-\tilde{Q}%
_{i}^{B}$, $W_{i}^{B}-\tilde{W}_{i}^{B}$, $f_{i}-\tilde{f}_{i}$ and $g_{i}-%
\tilde{g}_{i}$. Let $r_{i}$, $\hat{r}_{i}$, $s_{i}$, $\hat{s}_{i}$ be
defined as in the proof of Theorem \ref{linth1}, corresponding to $U-\tilde{U%
}$. We show that these quantities have upper bounds in the form of the right
hand side of (\ref{conth1.1}) in $D_{\delta }^{C}$, $D_{\delta }^{L}$ and $%
D_{\delta }^{R}$.

In $D_{\delta }^{C}$, (\ref{inteq1}) and (\ref{inteq3}) hold. Notice that
the functions $F_{i}^{R}$ and $F_{i}^{L}$ are linear in $r_{i}$, and $s_{i}$%
. Hence, there exists a constant $M$ (we will use $M$ generically for any
constant bounds that are independent of solutions) such that
\[
R_{i}^{C}\left( t\right) +S_{i}^{C}\left( t\right) \leq \left|
r_{i}^{I}\right| _{C\left[ 0,1\right] }+\left| s_{i}^{I}\right| _{C\left[
0,1\right] }+M\int_{0}^{t}\left( R_{i}^{C}\left( t^{\prime }\right)
+S_{i}^{C}\left( t^{\prime }\right) +T_{i}^{C}\left( t^{\prime }\right)
\right) dt^{\prime },
\]
where
\begin{equation}
R_{i}^{C}\left( t\right) =\sup_{\left\{ x:\left( x,t\right) \in D_{\delta
}^{C}\right\} }\left| r_{i}\left( x,t\right) \right| ,\quad S_{i}^{C}\left(
t\right) =\sup_{\left\{ x:\left( x,t\right) \in D_{\delta }^{C}\right\}
}\left| s_{i}\left( x,t\right) \right| ,  \label{conth1.3}
\end{equation}
and
\begin{equation}
T_{i}^{C}\left( t\right) =\sup_{\left\{ x:\left( x,t\right) \in D_{\delta
}^{C}\right\} }\left( \left| f_{i}\left( x,t\right) -\tilde{f}_{i}\left(
x,t\right) \right| +\left| g_{i}\left( x,t\right) -\tilde{g}_{i}\left(
x,t\right) \right| \right) .  \label{conth1.4}
\end{equation}
Hence, by Gronwall's inequality (see, e.g. \cite[p.327]{Mco95}),
\[
R_{i}^{C}\left( t\right) +S_{i}^{C}\left( t\right) \leq M\left( \left|
r_{i}^{I}\right| _{C\left[ 0,1\right] }+\left| s_{i}^{I}\right| _{C\left[
0,1\right] }+\delta \sup_{t\in \left( 0,\delta \right) }T_{i}^{C}\left(
t\right) \right)
\]
for $t\in \left[ 0,\delta \right] $. This proves that $R_{i}^{C}$ and $%
S_{i}^{C}$ have upper bounds in the form of the right side of (\ref{conth1.1}%
).

In $D_{\delta }^{L}$, if the left end is a source, we use either (\ref
{inteq5}) or (\ref{inteq5'}) according to the type of the boundary
condition. The resulting inequality has the form
\[
R_{i}^{L}\left( t\right) +\hat{S}_{i}^{L}\left( t\right) \leq \sigma \hat{S}%
_{i}^{L}\left( t\right) +M\left( \left| s_{i}^{I}\right| _{C\left[
0,1\right] }+\left| \xi _{i}^{B}\right| _{C\left[ 0,\delta \right]
}+\int_{0}^{t}\left( R_{i}^{L}\left( \tau \right) +\hat{S}_{i}^{L}\left(
\tau \right) +T_{i}^{L}\left( \tau \right) \right) d\tau \right)
\]
where $\xi _{i}^{B}$ is either $P_{i}^{B}$ or $Q_{i}^{B}$ depending on the
boundary condition, and $R_{i}^{L}$, $\hat{S}_{i}^{L}$ and $T_{i}^{L}$ are
defined in the same way as in (\ref{conth1.3})--(\ref{conth1.4}), with $%
D_{\delta }^{C}$ substituted by $D_{\delta }^{L}\cup D_{\delta }^{C}$, and $%
\sigma >0$ is a positive constant such that $\sigma =\varepsilon $ if the
boundary condition is (\ref{bcsp}) and
\[
\sigma =\varepsilon \sup_{t\in \left( 0,\delta \right) }\left| \frac{\lambda
_{i}^{L}\left( 0,t\right) }{\lambda _{i}^{R}\left( 0,t\right) }\right| <1
\]
if the boundary condition is (\ref{bcsq}). Replacing $M$ by $\left( 1-\sigma
\right) M$, we can write
\[
R_{i}^{L}\left( t\right) +\hat{S}_{i}^{L}\left( t\right) \leq M\left( \left|
s_{i}^{I}\right| _{C\left[ 0,1\right] }+\left| \xi _{i}^{B}\right| _{C\left[
0,\delta \right] }+\int_{0}^{t}\left( R_{i}^{L}\left( \tau \right) +\hat{S}%
_{i}^{L}\left( \tau \right) +T_{i}^{L}\left( \tau \right) \right) d\tau
\right) .
\]
Hence, by Gronwall's inequality
\[
R_{i}^{L}\left( t\right) +\hat{S}_{i}^{L}\left( t\right) \leq M\left( \left|
s_{i}^{I}\right| _{C\left[ 0,1\right] }+\left| \xi _{i}^{B}\right| _{C\left[
0,\delta \right] }+\delta \max_{t\in \left( 0,\delta \right)
}T_{i}^{L}\left( t\right) \right) .
\]
This proves that both $R_{i}^{L}\left( t\right) $ and $S_{i}^{L}\left(
t\right) $ have upper bounds in the form of the right hand side of (\ref
{conth1.1}).

If the left end is a junction, the solutions on the branches $j_{1},\ldots
,j_{\mu }$ connecting to the junction constitute a fixed point of the
operator $K$, which is defined in (\ref{linth1.8}). Let
\[
W\left( t\right) =\sum_{l=1}^{\nu }\left( \hat{R}_{j_{l}}^{R}\left( t\right)
+S_{j_{l}}^{R}\left( t\right) \right) +\sum_{l^{\prime }=\nu +1}^{\mu
}\left( R_{j_{l^{\prime }}}^{L}\left( t\right) +\hat{S}_{j_{l^{\prime
}}}^{L}\left( t\right) \right)
\]
where $\hat{R}_{i}^{R}$ and $S_{i}^{R}$ are defined as in (\ref{conth1.3})
with $D_{\delta }^{C}$ substituted by $D_{\delta }^{C}\cup D_{\delta }^{R}$.
Then, from $w=Kw$, we can deduce
\begin{eqnarray*}
W\left( t\right) &\leq &\sigma \left( \sum_{l=1}^{\nu }\hat{R}%
_{j_{l}}^{R}\left( t\right) +\sum_{l^{\prime }=\nu +1}^{\mu }\hat{S}%
_{j_{l^{\prime }}}^{L}\left( t\right) \right) \\
&&+M\left( \sum_{l=1}^{\nu }\left| r_{j_{l}}^{I}\right| _{C\left[ 0,1\right]
}+\sum_{l^{\prime }=\nu }^{\mu }\left| s_{j_{l^{\prime }}}^{I}\right|
_{C\left[ 0,1\right] }+\int_{0}^{t}\left( W\left( \tau \right) +T\left( \tau
\right) \right) d\tau \right) ,
\end{eqnarray*}
where
\[
T\left( \tau \right) =\sum_{l=1}^{\nu }T_{j_{l}}^{R}\left( \tau \right)
+\sum_{l^{\prime }=\nu +1}^{\mu }T_{j_{l^{\prime }}}^{L}\left( \tau \right)
\]
and $T_{i}^{R}\left( t\right) $ is defined as in (\ref{conth1.4}) with $%
D_{\delta }^{C}$ substituted by $D_{\delta }^{C}\cup D_{\delta }^{R}$.
Replacing $M$ by $\left( 1-\sigma \right) M$, we obtain
\[
W\left( t\right) \leq M\left( \sum_{l=1}^{\nu }\left| r_{j_{l}}^{I}\right|
_{C\left[ 0,1\right] }+\sum_{l^{\prime }=\nu }^{\mu }\left| s_{j_{l^{\prime
}}}^{I}\right| _{C\left[ 0,1\right] }+\int_{0}^{t}\left( W\left( \tau
\right) +T\left( \tau \right) \right) d\tau \right) .
\]
Hence, by Gronwall's inequality,
\[
W\left( t\right) \leq M\left( \sum_{l=1}^{\nu }\left| r_{j_{l}}^{I}\right|
_{C\left[ 0,1\right] }+\sum_{l^{\prime }=\nu }^{\mu }\left| s_{j_{l^{\prime
}}}^{I}\right| _{C\left[ 0,1\right] }+\delta \max_{t\in \left( 0,\delta
\right) }T\left( t\right) \right) .
\]
This leads to an upper bound in the form of the right hand side of (\ref
{conth1.1}) for $R_{i}^{R}\left( t\right) $, $S_{i}^{R}\left( t\right) $, $%
i=j_{1},\ldots ,j_{\nu }$, and $R_{i}^{L}\left( t\right) $, $S_{i}^{L}\left(
t\right) $, $i=j_{\nu +1},\ldots ,j_{\mu }$.

The only remaining case is when the right end of the branch is a terminal.
The fixed point equation to be used is either (\ref{inteq8}), (\ref{inteq8'}%
) or (\ref{inteq8''}) depending on the type of the boundary condition. In
the former two cases, the treatment is similar to that for sources. Hence,
we only consider the third case. From (\ref{inteq8''}), we obtain
\[
\hat{R}_{i}^{R}\left( t\right) +S_{i}^{R}\left( t\right) \leq \sigma \hat{R}%
_{i}^{R}\left( t\right) +M\left( \left| r_{i}^{I}\right| _{C\left[
0,1\right] }+\int_{0}^{t}\left( \hat{R}_{i}^{R}\left( t^{\prime }\right)
+S_{i}^{R}\left( t^{\prime }\right) +\left| W_{i}^{B}\left( t^{\prime
}\right) \right| +T_{i}^{R}\left( t^{\prime }\right) \right) dt^{\prime
}\right)
\]
where
\[
\sigma =\varepsilon \max_{t\in \left[ 0,\delta \right] }\left| \frac{%
m_{i}\left( t\right) }{n_{i}\left( t\right) }\right| <1.
\]
Hence, by Gronwall's inequality,
\[
\hat{R}_{i}^{R}\left( t\right) +S_{i}^{R}\left( t\right) \leq M\left( \left|
r_{i}^{I}\right| _{C\left[ 0,1\right] }+\delta \max_{t\in \left( 0,\delta
\right) }T_{i}^{R}\left( t\right) +\delta \max_{t\in \left( 0,\delta \right)
}\left| W_{i}^{B}\left( t\right) \right| \right) ,
\]
which gives the desired upper bounds of $R_{i}^{R}$ and $S_{i}^{R}$.

We have thus obtained an upper bound in the form of the right hand side of (%
\ref{conth1.1}) for the quantities $\left| r_{i}-\tilde{r}_{i}\right|
_{C\left( D_{\delta }\right) }$ and $\left| s_{i}-s_{i}\right| _{C\left(
D_{\delta }\right) }$. The conclusion of the lemma follows now from (\ref
{linth1.1}). \endproof

\section{The quasilinear system}

\setcounter{equation}{0} \setcounter{theorem}{0}In this section, we study
the quasilinear system where the coefficients $a_{i}$, $b_{i}$, $c_{i}$, $%
f_{i}$ and $g_{i}$ depend on both $\left( x,t\right) $ and $\left(
P_{i},Q_{i}\right) $. Under certain conditions, we show that the system has
a unique local solution. We then present a theorem on the continuity of
dependence of the solution on initial, boundary and forcing function.

The basic idea in the proof of the existence of solution is to construct an
iterative sequence. Substituting any vector function $\left(
p_{i},q_{i}\right) $ for $\left( P_{i},Q_{i}\right) $ in $a_{i}$, etc., the
system becomes linear. Thus, we can use Theorem \ref{linth1} to get a
solution $\left( P_{i},Q_{i}\right) $. This defines a mapping $S$ from $%
u=:\left( p_{i},q_{i}\right) $ to $U=:\left( P_{i},Q_{i}\right) $, and the
solution for the quasilinear system is a fixed point of $S$. If there is a
subset of a Banach space that is invariant under $S$, then, we can construct
a sequence
\[
u_{k+1}=Su_{k},\quad k=0,1,\ldots .
\]
In the case where the limit exists and is unique, it gives rise to fixed
point of $S$. This is our approach in this section.

In this approach, conditions (\ref{cond2}) and (\ref{cond3}) are repeatedly
used. One might want to impose them for all the values of the variables.
This would give the existence and uniqueness for the global solution, as in
the case of the linear system. However, such a requirement is so restrictive
that even the original system (\ref{deaq}) cannot meet it. Therefore, we
will impose them only for $t=0$, and obtain the local solution for the
quasilinear system.

\begin{theorem}
\label{quath1}Assume that the initial and boundary functions $P_{i}^{I}$, $%
Q_{i}^{I}$, $P_{i}^{B}$, $Q_{i}^{B}$, $W_{i}^{B}$ and the system functions $%
a_{i}$, $b_{i}$, $c_{i}$, $f_{i}$, $g_{i}$ all have continuous first-order
derivatives with respect to each variable. Suppose that $a_{i}>0$ for all
the values of its arguments, and that conditions (\ref{cond2})--(\ref{cond3}%
) hold at $t=0$. Suppose also that the initial functions $P_{i}^{I}$, $%
Q_{i}^{I}$ satisfy any relevant boundary conditions at $t=0$. Then, for some
$\delta >0$, there is a unique solution for $0\leq t<\delta $ to the
quasilinear system (\ref{depq}) with the initial and boundary conditions
described in Section \ref{Int}.
\end{theorem}

\paragraph{\noindent Proof.}

We first consider the simpler case where $U^{I}=:\left( P^{I},Q^{I}\right)
=0 $. Let $v=\left\{ v_{i}\right\} $, $v_{i}=\left( p_{i},q_{i}\right) $ be
a family of vector functions (not necessarily constitutes a solution) that
satisfy the initial and boundary conditions. Substitute $v$ for $U$ in the
functions $a_{i}$, $b_{i}$, $c_{i}$, $f_{i}$ and $g_{i}$. Then, the system
becomes linear and we can invoke Theorem \ref{linth1} to obtain a solution $%
U $ to the linear system. This defines a mapping $S:$ $v\mapsto U$. A
solution of the quasilinear system is then a fixed point of $S$. We will
choose a subset $X_{\delta ,M_{0}}$ of a Banach space such that (1) $%
SX_{\delta ,M_{0}}\subset X_{\delta ,M_{0}}$, and (2) $S$ is contracting in $%
X_{\delta ,M_{0}}$. For any scalar or vector function $f\in C^{k}\left(
D_{\delta }\right) $, let $\left| f\right| _{k,\delta }$ denote the maximum
norm of all the $k$-th order derivatives of $f$ in $D_{\delta }$. (If $f$ is
a vector function, $\left| f\right| _{k,\delta }=\max_{i}\left\{ \left|
f_{i}\right| _{k,\delta }\right\} $.) Let $C_{B}\left( D_{\delta },\Bbb{R}%
^{2n}\right) $ denote the subset of the vector-valued functions in $C\left(
D_{\delta },\Bbb{R}^{2n}\right) $ that satisfy the initial and boundary
conditions. We seek $X_{\delta ,M_{0}}$ in the form
\begin{equation}
X_{\delta ,M_{0}}=\left\{ v\in C_{B}\left( D_{\delta },\Bbb{R}^{2n}\right) :%
\text{ }\left| v\right| _{0,\delta }\leq M_{0},\left| v\right| _{1,\delta
}\leq M_{1}\right\}  \label{quath1.5}
\end{equation}
where $M_{0}$ is an arbitrary positive constant and $M_{1}$ is a constant to
be determined. Note that by the vanishing initial condition, for any $M_{1}$%
, $\left| U\right| _{1,\delta }\leq M_{1}$ implies $\left| U\right|
_{0,\delta }\leq M_{1}\delta $. Hence, for any $M_{0}$, we can ensure $%
\left| U\right| _{0,\delta }\leq M_{0}$ by reducing $\delta $. It remains,
therefore, only to show that for $M_{1}$ sufficiently large and $\delta $
sufficiently small, $\left| v\right| _{1,\delta }\leq M_{1}$ implies $\left|
Sv\right| _{1,\delta }\leq M_{1}$. Throughout this proof, we use $M$ to
represent any positive constant that may depend on $M_{1}$ but is otherwise
independent of $v$ and $\delta $, and use $\tilde{M}$ for any constant that
is independent of $M_{1}$, $v$ and $\delta $. The values of $M$ or $\tilde{M}
$ in different occurrences need not be equal.

Let $U=Sv$ and let $r_{i}$ and $s_{i}$ be defined by (\ref{linth1.6}). On
each branch, we show that
\begin{equation}
\max \left\{ \left| \left( r_{i}\right) _{x}\right| ,\left| \left(
s_{i}\right) _{x}\right| ,\right\} \leq M_{1}  \label{quath1.11}
\end{equation}
and
\begin{equation}
\max \left\{ \left| \left( r_{i}\right) _{t}\right| ,\left| \left(
s_{i}\right) _{t}\right| \right\} \leq M_{1}  \label{quath1.12}
\end{equation}
in $D_{\delta }^{C}$, $D_{\delta }^{L}$ and $D_{\delta }^{R}$ if $M_{1}$ is
large and $\delta $ is small. (Recall that $D_{\delta }^{C}$ etc. are the
intersections $D_{i}^{C}\cap D_{\delta }$ etc., respectively.) In fact, only
(\ref{quath1.11}) needs to be shown. To see this, first observe that the
vanishing initial condition and the compatibility of the initial and
boundary conditions gives
\[
\max_{i}\left\{ \left| P_{i}^{B}\right| _{C\left[ 0,\delta \right] },\left|
Q_{i}^{B}\right| _{C\left[ 0,\delta \right] }\right\} \leq M\delta .
\]
Hence, we obtain from Lemma \ref{conth1} with $\tilde{U}=0$ that
\begin{equation}
\left| U\right| _{0,\delta }\leq M\delta .  \label{quath1.1}
\end{equation}
From (\ref{linth1.2}) and (\ref{linth1.3}), there are constants $\tilde{M}$
and $M$ such that
\begin{equation}
\begin{array}{l}
\left| \partial _{i}^{R}r_{i}\right| \leq \left| l_{i}^{R}F_{i}\right|
+\left| \partial _{i}^{R}l_{i}^{R}\right| \left| U_{i}\right| \leq \tilde{M}%
+M\delta , \\[12pt]
\left| \partial _{i}^{L}s_{i}\right| \leq \left| l_{i}^{L}F_{i}\right|
+\left| \partial _{i}^{L}l_{i}^{L}\right| \left| U_{i}\right| \leq \tilde{M}%
+M\delta
\end{array}
\label{quath1.10}
\end{equation}
for each $i=1,\ldots ,n$. Hence, (\ref{quath1.12}) follows from (\ref
{quath1.11}), (\ref{quath1.10}) and the definition of $\partial _{i}^{L}$
and $\partial _{i}^{R}$ in (\ref{linth1.5}). We also note that (\ref
{linth1.1}) and (\ref{quath1.10}) imply
\begin{equation}
\left| \partial _{i}^{R}U_{i}\right| _{0,\delta }\leq \tilde{M}+M\delta
,\quad \left| \partial _{i}^{R}U_{i}\right| _{0,\delta }\leq \tilde{M}%
+M\delta  \label{quath1.7}
\end{equation}
for all $i$. This will be used later.

We first consider the middle region $D_{\delta }^{C}$, where the solution $%
\left( r_{i},s_{i}\right) $ satisfies the integral equations (\ref{inteq1})
and (\ref{inteq3}) with $r_{i}^{I}=s_{i}^{I}=0$. Differentiating the
equations with respect to $x$, we have
\begin{equation}
\begin{array}{l}
\displaystyle\left( r_{i}\right) _{x}=\left( l_{i}^{R}\right)
_{x}U_{i}\left( x,t\right) +\int_{0}^{t}\left[ \left( l_{i}^{R}F_{i}\right)
_{x}+\left( \partial _{i}^{R}l_{i}^{R}\right) \left( U_{i}\right)
_{x}-\left( l_{i}^{R}\right) _{x}\left( \partial _{i}^{R}U_{i}\right)
\right] \left( x_{i}^{R}\right) _{x}dt, \\
\displaystyle\left( s_{i}\right) _{x}=\left( l_{i}^{L}\right)
_{x}U_{i}\left( x,t\right) +\int_{0}^{t}\left[ \left( l_{i}^{L}F_{i}\right)
_{x}+\left( \partial _{i}^{L}l_{i}^{L}\right) \left( U_{i}\right)
_{x}-\left( l_{i}^{L}\right) _{x}\left( \partial _{i}^{L}U_{i}\right)
\right] \left( x_{i}^{L}\right) _{x}dt.
\end{array}
\label{quath1.2}
\end{equation}
Here, we used an identity from \cite[p.469]{CH62}:
\begin{equation}
\begin{array}{r}
\displaystyle\frac{d}{d\xi }\int_{a}^{b}f\left( x\left( t\right) ,t\right)
Dg\left( x\left( t\right) ,t\right) dt=f\left( x\left( b\right) ,b\right)
g_{x}\left( x\left( b\right) ,b\right) -f\left( x\left( a\right) ,a\right)
g_{x}\left( x\left( a\right) ,a\right) \\
\displaystyle+\int_{a}^{b}\left[ f_{x}\left( x\left( t\right) ,t\right)
Dg\left( x\left( t\right) ,t\right) -Df\left( x\left( t\right) ,t\right)
g_{x}\left( x\left( t\right) ,t\right) \right] dt
\end{array}
\label{quath1.8}
\end{equation}
where $x\left( t\right) $ is a function such that $x\left( b\right) =\xi $
and $D=\frac{\partial }{\partial t}+x^{\prime }\left( t\right) \frac{%
\partial }{\partial x}$. Let
\begin{equation}
R_{i}^{C}\left( t\right) =\sup_{\left\{ x:\left( x,t\right) \in D_{\delta
}^{C}\right\} }\left\{ \left| \left( r_{i}\right) _{x}\left( x,t\right)
\right| \right\} ,\quad S_{i}^{C}\left( t\right) =\sup_{\left\{ x:\left(
x,t\right) \in D_{\delta }^{C}\right\} }\left\{ \left| \left( s_{i}\right)
_{x}\left( x,t\right) \right| \right\} .  \label{quath1.9}
\end{equation}
From (\ref{quath1.1}), (\ref{quath1.7}) and (\ref{quath1.2}), we derive
\[
R_{i}^{C}\left( t\right) +S_{i}^{C}\left( t\right) \leq M\delta
+M\int_{0}^{t}\left( 1+R_{i}^{C}\left( t^{\prime }\right) +S_{i}^{C}\left(
t^{\prime }\right) \right) dt^{\prime }
\]
for $t\in \left[ 0,\delta \right] $. Hence, Gronwall's inequality gives
\[
\left| \left( r_{i}\right) _{x}\right| \leq M\delta e^{M\delta },\quad
\left| \left( s_{i}\right) _{x}\right| \leq M\delta e^{M\delta }
\]
in $D_{\delta }^{C}$. This proves (\ref{quath1.11}) in $D_{\delta }^{C}$ if $%
M_{1}$ is sufficiently large and $\delta $ is sufficiently small$.$

We next consider the left triangular region $D_{\delta }^{L}$ in the case
where the branch is connected to a source. Let $\hat{s}_{i}=s_{i}/%
\varepsilon $ for any $\varepsilon >0$. Then, the pair $\left( r_{i},\hat{s}%
_{i}\right) $ satisfies the fixed point equations of either (\ref{inteq5})
or (\ref{inteq5'}), depending on the type of the boundary condition.
Differentiating the equations with respect to $x$ and using a slightly
modified version of (\ref{quath1.8}), we have
\begin{equation}
\begin{array}{l}
\displaystyle\left( r_{i}\right) _{x}=\left( \zeta
_{i}-l_{i}^{R}F_{i}-\left( \partial _{i}^{R}l_{i}^{R}\right) U_{i}-\left(
l_{i}^{R}\right) _{x}U_{i}\right) \left( 0,\tau \right) \tau _{x}+\left(
l_{i}^{R}\right) _{x}U_{i}\left( x,t\right) \\
\displaystyle\hspace{0.5in}+\int_{\tau }^{t}\left[ \left(
l_{i}^{R}F_{i}\right) _{x}+\left( \partial _{i}^{R}l_{i}^{R}\right) \left(
U_{i}\right) _{x}-\left( l_{i}^{R}\right) _{x}\left( \partial
_{i}^{R}U_{i}\right) \right] \left( x_{i}^{R}\right) _{x}dt, \\
\displaystyle\left( \hat{s}_{i}\right) _{x}=\frac{1}{\varepsilon }\left(
l_{i}^{L}\right) _{x}U_{i}\left( t,x\right) +\frac{1}{\varepsilon }%
\int_{0}^{t}\left[ \left( l_{i}^{L}F_{i}\right) _{x}+\left( \partial
_{i}^{L}l_{i}^{L}\right) \left( U_{i}\right) _{x}-\left( l_{i}^{L}\right)
_{x}\left( \partial _{i}^{L}U_{i}\right) \right] \left( x_{i}^{L}\right)
_{x}dt,
\end{array}
\label{quath1.15}
\end{equation}
where
\[
\zeta _{i}=2\left( u_{i}P_{i}^{B}\right) _{t}+\varepsilon \left( \hat{s}%
_{i}\right) _{t}
\]
if the boundary condition is given by (\ref{bcsp}), and
\[
\zeta _{i}=2\left( \frac{a_{i}u_{i}}{\lambda _{i}^{R}}Q_{i}^{B}\right)
_{t}+\varepsilon \left( \frac{\lambda _{i}^{L}}{\lambda _{i}^{R}}\right) _{t}%
\hat{s}_{i}+\varepsilon \left( \frac{\lambda _{i}^{L}}{\lambda _{i}^{R}}%
\right) \left( \hat{s}_{i}\right) _{t}
\]
if the boundary condition is given by (\ref{bcsq}). (Modification of (\ref
{quath1.8}) is caused by the lower limit of the integral in the first
equation of (\ref{quath1.15}) which also depends on $x$.) This equation is
valid for any $\varepsilon $. So, we may choose $\varepsilon $ so small such
that
\[
\sigma =:\varepsilon \left| \lambda _{i}^{L}\tau _{x}\left( 0,t\right)
\right| \max \left\{ 1,\left| \left( \frac{\lambda _{i}^{L}\left( 0,t\right)
}{\lambda _{i}^{R}\left( 0,t\right) }\right) \right| \right\} <1,\quad t\in
\left[ 0,\delta \right] .
\]
To proceed further, we need an estimate of $\left| \tau _{x}\left(
0,t\right) \right| $. Observe that $\tau \left( x\right) $ satisfies the
equation
\[
x_{i}^{R}\left( \tau ;x,t\right) =0
\]
where $x_{i}^{R}\left( \tau ;x,t\right) $ is the solution of the initial
value problem
\[
\frac{dx_{i}^{R}}{ds}=\lambda _{i}^{R}\left( x_{i}^{R},s\right) ,\quad
x_{i}^{R}\left( t;x,t\right) =x.
\]
By differentiation,
\begin{equation}
\lambda _{i}^{R}\left( 0,\tau \left( x\right) \right) \tau _{x}+\left. \frac{%
\partial x_{i}^{R}}{\partial x}\right| _{\left( \tau \left( x\right)
;x,t\right) }=0.  \label{quath1.16}
\end{equation}
Let $w_{i}=\partial x_{i}^{R}/\partial x$. Then, $w_{i}$ is the solution of
the linear equation
\[
\frac{dw_{i}}{ds}=\left( \lambda _{i}^{R}\right) _{x}\left( x_{i}^{R}\left(
s;x,t\right) ,s\right) w_{i},\quad w_{i}\left( t\right) =1.
\]
Solving the equation,
\[
w_{i}\left( s\right) =\exp \left( \int_{t}^{s}\left( \lambda _{i}^{R}\right)
_{x}\left( x_{i}^{R}\left( s^{\prime };x,t\right) ,s^{\prime }\right)
ds^{\prime }\right) .
\]
Returning to (\ref{quath1.16}), we find
\[
\tau _{x}=\frac{-1}{\lambda _{i}^{R}\left( 0,\tau \left( x\right) \right) }%
\exp \left( \int_{t}^{\tau \left( x\right) }\left( \lambda _{i}^{R}\right)
_{x}\left( x_{i}^{R}\left( s^{\prime };x,t\right) ,s^{\prime }\right)
ds^{\prime }\right) .
\]
Observe that $0<\tau \left( x\right) <t\leq \delta $ and the integrand is
bounded. Hence,
\begin{equation}
\left| \tau _{x}\right| \leq \tilde{M}e^{M\delta }.  \label{quath1.13}
\end{equation}
This is the estimate we need. By this estimate, for any $M_{1}$, we can
choose $\delta $ small enough such that the constants $\sigma $ and $%
\varepsilon $ are independent of $M_{1}$. Let $R_{i}^{L}\left( t\right) $
and $\hat{S}_{i}^{L}\left( t\right) $ be defined as in (\ref{quath1.9})
except that $s_{i}$ is substituted by $\hat{s}_{i}$ and $D_{\delta }^{C}$ is
substituted by $D_{\delta }^{L}\cup D_{\delta }^{C}$, We derive from (\ref
{quath1.15}) and the identity
\[
\left( \hat{s}_{i}\right) _{t}=\partial _{i}^{L}\hat{s}_{i}-\lambda
_{i}^{L}\left( \hat{s}_{i}\right) _{x}
\]
that
\[
R_{i}^{L}\left( t\right) +\hat{S}_{i}^{L}\left( t\right) \leq \sigma \hat{S}%
_{i}^{L}\left( t\right) +\tilde{M}+M\delta +M\int_{0}^{t}\left(
1+R_{i}^{L}\left( t^{\prime }\right) +\hat{S}_{i}^{L}\left( t^{\prime
}\right) \right) dt^{\prime }.
\]
Replacing $M$ and $\tilde{M}$ by $M\left( 1-\sigma \right) $ and $\tilde{M}%
\left( 1-\sigma \right) $, respectively, and applying Gronwall's inequality,
we obtain
\[
R_{i}^{L}\left( t\right) +\hat{S}_{i}^{L}\left( t\right) \leq \left( \tilde{M%
}+M\delta \right) e^{M\delta }.
\]
Since $\left| s_{i}\right| \leq \left| \hat{s}_{i}\right| $, it follows that
\[
\max \left\{ \left| \left( r_{i}\right) _{x}\right| ,\left| \left(
s_{i}\right) _{x}\right| \right\} \leq \left( \tilde{M}+M\delta \right)
e^{M\delta }
\]
in $D_{\delta }^{L}\cup D_{\delta }^{C}$. This proves (\ref{quath1.11}) in $%
D_{\delta }^{L}\cup D_{\delta }^{C}$ if $M_{1}$ is large and $\delta $ is
small.

We next consider the case where the left end of the branch is a junction. As
before, we shall consider the branches that are connected to the same
junction simultaneously. This also includes the right triangular regions $%
D_{\delta }^{R}$ for the branches that are connected to the junction from
left. We consider the fixed point equation $w=Kw$ where $w$ and $Kw$ are
defined in (\ref{linth1.7}) and (\ref{linth1.8}), respectively.
Differentiating the equations, we obtain (\ref{quath1.15}) in $D_{\delta
}^{L}\cup D_{\delta }^{C}$ for $i=j_{\nu +1},\ldots ,j_{\mu }$ and
\begin{equation}
\begin{array}{l}
\displaystyle\left( \hat{r}_{i}\right) _{x}=\frac{1}{\varepsilon }\left(
l_{i}^{R}\right) _{x}U_{i}\left( t,x\right) +\frac{1}{\varepsilon }%
\int_{0}^{t}\left[ \left( l_{i}^{R}F_{i}\right) _{x}+\left( \partial
_{i}^{R}l_{i}^{R}\right) \left( U_{i}\right) _{x}-\left( l_{i}^{R}\right)
_{x}\left( \partial _{i}^{R}U_{i}\right) \right] \left( x_{i}^{R}\right)
_{x}dt, \\[12pt]
\displaystyle\left( s_{i}\right) _{x}=\left( \theta
_{i}-l_{i}^{L}F_{i}-\left( \partial _{i}^{L}l_{i}^{L}\right) U_{i}-\left(
l_{i}^{L}\right) _{x}U_{i}\right) \left( 1,\tau \right) \tau _{x}+\left(
l_{i}^{L}\right) _{x}U_{i}\left( x,t\right) \\
\displaystyle\hspace{0.5in}+\int_{\tau }^{t}\left[ \left(
l_{i}^{L}F_{i}\right) _{x}+\left( \partial _{i}^{L}l_{i}^{L}\right) \left(
U_{i}\right) _{x}-\left( l_{i}^{L}\right) _{x}\left( \partial
_{i}^{L}U_{i}\right) \right] \left( x_{i}^{L}\right) _{x}dt,
\end{array}
\label{quath1.14}
\end{equation}
in $D_{\delta }^{C}\cup D_{\delta }^{R}$ for $i=j_{1},\ldots ,j_{\nu }$,
where
\[
\begin{array}{l}
\displaystyle\zeta _{i}=\varepsilon \sum_{l=1}^{\nu }\left( \left(
n_{j_{l}}^{i}\right) _{t}\hat{r}_{j_{l}}\left( 1,\tau \right)
+n_{j_{l}}^{i}\left( \hat{r}_{j_{l}}\right) _{t}\right) +\varepsilon
\sum_{l^{\prime }=\nu +1}^{\mu }\left( \left( n_{j_{l^{\prime }}}^{i}\right)
_{t}\hat{s}_{j_{l^{\prime }}}\left( 0,\tau \right) +n_{j_{l^{\prime
}}}^{i}\left( \hat{s}_{j_{l^{\prime }}}\right) _{t}\left( 0,\tau \right)
\right) , \\
\displaystyle\theta _{i}=\varepsilon \sum_{l=1}^{\nu }\left( \left(
m_{j_{l}}^{i}\right) _{t}\hat{r}_{j_{l}}\left( 1,\tau \right)
+m_{j_{l}}^{i}\left( \hat{r}_{j_{l}}\right) _{t}\right) +\varepsilon
\sum_{l^{\prime }=\nu +1}^{\mu }\left( \left( m_{j_{l^{\prime }}}^{i}\right)
_{t}\hat{s}_{j_{l^{\prime }}}\left( 0,\tau \right) +m_{j_{l^{\prime
}}}^{i}\left( \hat{s}_{j_{l^{\prime }}}\right) _{t}\left( 0,\tau \right)
\right) ,
\end{array}
\]
and $m_{j}^{i}$, $n_{j}^{i}$ are defined in the proof of Theorem \ref{linth1}%
. Note that the estimate (\ref{quath1.13}) holds for $\tau _{x}$ in both (%
\ref{quath1.15}) and (\ref{quath1.14}), although in the latter case, $\tau $
is the $t$-coordinate of the intersection of the left-going characteristic
curve $x_{i}^{L}$ with the vertical line $x=1$. The derivation is identical.
Hence, there is a constant $\varepsilon $, independent of $M_{1}$, such that
\begin{eqnarray*}
\varepsilon \left| \tau _{x}\right| \left( \sum_{k=1}^{\nu }\left|
m_{j_{k}}^{i}\left( t\right) \right| +\sum_{k^{\prime }=\nu +1}^{\mu }\left|
m_{j_{k^{\prime }}}^{i}\left( t\right) \right| \right) &<&1, \\
\varepsilon \left| \tau _{x}\right| \left( \sum_{k=1}^{\nu }\left|
n_{j_{k}}^{i}\left( t\right) \right| +\sum_{k^{\prime }=\nu +1}^{\mu }\left|
n_{j_{k^{\prime }}}^{i}\left( t\right) \right| \right) &<&1
\end{eqnarray*}
in $\left[ 0,\delta \right] $. Let $\sigma $ be the maximum of the
quantities on the left hand side of the above inequalities. Define $\hat{R}%
_{i}^{R}$, $S_{i}^{R}$, $R_{i}^{L}$ and $\hat{S}_{i}^{L}$ as in (\ref
{quath1.9}) with obvious modifications. We see that the function
\[
W\left( t\right) =\sum_{l=1}^{\nu }\left( \hat{R}_{j_{l}}^{R}\left( t\right)
+S_{j_{l}}^{R}\left( t\right) \right) +\sum_{l^{\prime }=\nu +1}^{\mu
}\left( R_{j_{l^{\prime }}}^{L}\left( t\right) +\hat{S}_{j_{l^{\prime
}}}^{L}\left( t\right) \right)
\]
satisfies the inequality
\begin{eqnarray*}
\left( 1-\sigma \right) W\left( t\right) &\leq &\sum_{l=1}^{\nu }\left(
\left( 1-\sigma \right) \hat{R}_{j_{l}}^{R}\left( t\right)
+S_{j_{l}}^{R}\left( t\right) \right) +\sum_{l^{\prime }=\nu +1}^{\mu
}\left( R_{j_{l^{\prime }}}^{L}\left( t\right) +\left( 1-\sigma \right) \hat{%
S}_{j_{l^{\prime }}}^{L}\left( t\right) \right) \\
&\leq &\tilde{M}+M\delta +M\int_{0}^{t}\left( 1+W\left( t^{\prime }\right)
\right) dt^{\prime }.
\end{eqnarray*}
Hence, by rescaling and using Gronwall's inequality, we achieve
\[
W\left( t\right) \leq \left( \tilde{M}+M\delta \right) e^{M\delta }.
\]
This proves that
\[
\max \left\{ \left| \left( r_{i}\right) _{x}\right| ,\left| \left(
s_{i}\right) _{x}\right| \right\} \leq M_{1}
\]
in $D_{\delta }^{R}$ for $i=j_{1},\ldots ,j_{\nu }$ and in $D_{\delta }^{L}$
for $i=j_{\nu +1},\ldots ,j_{\mu }$ if $M_{1}$ is sufficiently large and $%
\delta $ is sufficiently small. We have thus proved (\ref{quath1.11}) in
this case.

\smallskip It remains to treat the branches that are connected to terminals.
If the terminal boundary condition is either (\ref{bctp}) or (\ref{bctq}),
the argument is parallel to the one given above for sources. Hence, we only
consider the case where the boundary condition is (\ref{bctw}). The fixed
point equation in this case is (\ref{inteq8''}). Differentiating (\ref
{inteq8''}) with respect to $x$ gives (\ref{quath1.14}) with
\[
\zeta _{i}=\varepsilon \left( \frac{m_{i}}{n_{i}}\right) _{t}\tau _{x}\hat{r}%
_{i}\left( 1,\tau \right) +\varepsilon \frac{m_{i}}{n_{i}}\tau _{x}\left(
\hat{r}_{i}\right) _{t}\left( 1,\tau \right) -\left( \frac{1}{n_{i}}\right)
_{t}\int_{0}^{t}H_{i}\left( t^{\prime },r_{i}\left( 1,t^{\prime }\right)
,s_{i}\left( 1,t^{\prime }\right) \right) dt^{\prime }.
\]
Let $\delta $ be sufficiently small such that $\left| \tau _{x}\right| $ is
bounded by a constant independent of $M_{1}$. Choose $\varepsilon >0$ such
that
\[
\sigma =:\varepsilon \left| \lambda _{i}^{R}\left( 1,t\right) \right| \left|
\frac{m_{i}}{n_{i}}\tau _{x}\left( 1,t\right) \right| <1
\]
for $t\in \left[ 0,\delta \right] $. Note that $\left( \frac{m_{i}}{n_{i}}%
\right) _{t}$ and $\left( \frac{1}{n_{i}}\right) _{t}$ are bounded (by a
constant depending on $M_{1}$). Hence,
\[
\hat{R}_{i}^{R}\left( t\right) +S_{i}^{R}\left( t\right) \leq \sigma \hat{R}%
_{i}^{R}\left( t\right) +\tilde{M}+M\delta +M\int_{0}^{t}\left( 1+\hat{R}%
_{i}^{R}\left( t^{\prime }\right) +S_{i}^{R}\left( t^{\prime }\right)
\right) dt^{\prime }.
\]
This leads to
\[
\hat{R}_{i}^{R}\left( t\right) +S_{i}^{R}\left( t\right) \leq \left( \tilde{M%
}+M\delta \right) e^{M\delta }
\]
in $D_{\delta }^{R}$ upon rescaling of constants. Hence, (\ref{quath1.11})
holds in $D_{\delta }^{R}$.

This completes the proof of (\ref{quath1.11}) in all cases. By choosing
appropriate values of $M_{1}$ and $\delta $, we thus obtain a set $X_{\delta
,M_{0}}$ in the form of (\ref{quath1.5}) which is invariant under the
mapping $S$.

We now show that $S$ is a contraction in $X_{\delta ,M_{0}}$. Let $U=Sv$, $%
\tilde{U}=S\tilde{v}$ for some $v,\tilde{v}\in X_{\delta }$, and let $W=U-%
\tilde{U}$. $W$ satisfies the vanishing initial and external boundary
conditions and its differential equations takes the form of (\ref{depq})
with the coefficients
\[
a_{i}=a_{i}\left( x,t,v\right) ,\ b_{i}=b_{i}\left( x,t,v\right) ,\
c_{i}=c_{i}\left( x,t,v\right)
\]
and the forcing functions $f_{i}$ and $g_{i}$ replaced by
\begin{equation}
\hat{f}_{i}=:f_{i}\left( x,t,v\right) -f_{i}\left( x,t,\tilde{v}\right)
+\left( a_{i}\left( x,t,v\right) -a_{i}\left( x,t,\tilde{v}\right) \right)
\frac{\partial \tilde{Q}_{i}}{\partial x}  \label{quath1.17}
\end{equation}
and
\begin{equation}
\begin{array}{r}
\displaystyle\hat{g}_{i}=:g_{i}\left( x,t,v\right) -g_{i}\left( x,t,\tilde{v}%
\right) +\left( b_{i}\left( x,t,v\right) -b_{i}\left( x,t,\tilde{v}\right)
\right) \frac{\partial \tilde{P}_{i}}{\partial x} \\[12pt]
\displaystyle+2\left( c_{i}\left( x,t,v\right) -c_{i}\left( x,t,\tilde{v}%
\right) \right) \frac{\partial \tilde{Q}_{i}}{\partial x},
\end{array}
\label{quath1.18}
\end{equation}
respectively. By the Lipschitz property and the boundedness $\left| \tilde{U}%
\right| _{1,\delta }\leq M_{1}$, there is a constant $M$ such that
\[
\left| \hat{f}\right| _{0,\delta }\leq M\left| v-\tilde{v}\right| _{0,\delta
},\quad \left| \hat{g}\right| _{0,\delta }\leq M\left| v-\tilde{v}\right|
_{0,\delta }.
\]
Hence, by Theorem \ref{conth1},
\[
\left| Sv-S\tilde{v}\right| _{0,\delta }\leq M\delta \left| v-\tilde{v}%
\right| _{0,\delta }.
\]
Therefore, $S$ is contracting in $X_{\delta ,M_{0}}$ if $\delta $ is
sufficiently small.

The rest is standard (cf. e.g., \cite{CH62}). Starting with a $v_{0}\in
X_{\delta ,M_{0}}$, we generate an iterative sequence $v_{k+1}=Sv_{k}$.
Clearly, each $v_{k}$ lies in $X_{\delta ,M_{0}}$ and the sequence converges
uniformly. The limit then satisfies the integral equations in the proof of
Theorem \ref{linth1}, and hence, is differentiable. Therefore, it is the
solution of the quasilinear differential equations. This proves the
existence and uniqueness of the solution when $U^{I}=0$.

If $U^{I}\neq 0$, we regard $U^{I}$ as a vector function of $x$ and $t$ and
introduce $\tilde{U}=U-U^{I}$. It follows that $\tilde{U}$ is a solution of
the quasilinear equations (\ref{depq}) with the forcing functions $\tilde{f}%
_{i}$ and $\tilde{g}_{i}$ given by
\[
\tilde{f}_{i}=f_{i}-\left( Q_{i}^{I}\right) _{x}a_{i},\quad \tilde{g}%
_{i}=g_{i}-\left( P_{i}^{I}\right) _{x}b_{i}-\left( Q_{i}^{I}\right)
_{x}2c_{i}
\]
and the boundary functions are given by
\[
\tilde{P}_{i}^{B}=P_{i}^{B}-P_{i}^{I},\ \tilde{Q}%
_{i}^{B}=Q_{i}^{B}-Q_{i}^{I},\ \tilde{W}_{i}^{B}=W_{i}^{B}-\delta
_{i}P_{i}^{I}+\varepsilon _{i}Q_{i}^{I}.
\]
Since $\tilde{U}$ has the vanishing initial values, it can be uniquely
solved for an interval of $t\in \left[ 0,\delta \right] $. This gives rise
to a solution $U$. \endproof

\paragraph{\emph{Remark:}}

Examples can be constructed to show that if the condition (\ref{cond3})
fails at $t=0$, then, the local solution need not exist or may be not
unique. In particular, if (\ref{cond3}) fails at a source end, then, the
system is under-determined, and if it fails at a terminal end, the system is
over-determined. See Section \ref{conc} for further discussion.

We give next a result for the continuity of dependence of the solution and
its derivatives on the initial, boundary and forcing functions and their
derivatives. This follows from an argument similar to the proofs of Lemma
\ref{conth1} and Theorem \ref{quath1}.

\begin{corollary}
\label{quath2}Let $U=\left( P,Q\right) $ and $\tilde{U}=\left( \tilde{P},%
\tilde{Q}\right) $ be two solutions of the quasilinear problem of Theorem
\ref{quath1}. Suppose the conditions of that theorem hold for the initial
and boundary functions of both solutions. Then, there exists a constant $M>0$%
, independent of initial, boundary and forcing functions, such that
\begin{equation}
\begin{array}{r}
\left| U-\tilde{U}\right| _{k,\delta }\leq M\left( \left| P^{I}-\tilde{P}%
^{I}\right| _{C^{k}\left[ 0,1\right] }+\left| Q^{I}-\tilde{Q}^{I}\right|
_{C^{k}\left[ 0,1\right] }+\left| P^{B}-\tilde{P}^{B}\right| _{C^{k}\left[
0,\delta \right] }+\left| Q^{B}-\tilde{Q}^{B}\right| _{C^{k}\left[ 0,\delta
\right] }\right. \\
+\left. \delta \left| f-\tilde{f}\right| _{C^{k}\left( \overline{D_{\delta }}%
\right) }+\delta \left| g-\tilde{g}\right| _{C^{k}\left( \overline{D_{\delta
}}\right) }+\delta \left| W-\tilde{W}\right| _{C^{k}\left[ 0,\delta \right]
}\right) .
\end{array}
\label{quath2.1}
\end{equation}
for $k=0,1$.
\end{corollary}

\paragraph{Proof.}

For $k=0$, the result follows from substituting one of the solutions into
the coefficients, modifying the forcing functions by (\ref{quath1.17})--(\ref
{quath1.18}), and using Lemma \ref{conth1}. For $k=1$, we differentiate the
equations and apply the lemma to the resulting equations for the derivatives
of the solution. The process is standard and is omitted. \endproof

\section{A finite-difference scheme}

\setcounter{equation}{0} \setcounter{theorem}{0}In this section, we present
a finite-difference scheme that computes discretized solutions, and prove
the convergence of the scheme.

The scheme is based on the equations in (\ref{linth1.2}). Substituting (\ref
{linth1.6}) and (\ref{linth1.3}) into (\ref{linth1.2}), we obtain the normal
form of the equations
\[
\begin{array}{l}
-\lambda _{i}^{L}P_{i,t}+a_{i}Q_{i,t}+\lambda _{i}^{R}\left( -\lambda
_{i}^{L}P_{i,x}+a_{i}Q_{i,x}\right) =d_{i}^{R}, \\[12pt]
-\lambda _{i}^{R}P_{i,t}+a_{i}Q_{i,t}+\lambda _{i}^{L}\left( -\lambda
_{i}^{R}P_{i,x}+a_{i}Q_{i,x}\right) =d_{i}^{L},
\end{array}
\]
where
\[
d_{i}^{R}\left( x,t,P_{i},Q_{i}\right) =-\lambda
_{i}^{L}f_{i}+a_{i}g_{i},\quad d_{i}^{L}\left( x,t,P_{i},Q_{i}\right)
=-\lambda _{i}^{R}f_{i}+a_{i}g_{i}.
\]
Let $h$ and $k$ be the spatial and temporal step sizes, respectively. Hence,
$hN=1$ for some integer $N$. We impose the finite-difference equations as
\begin{equation}
\begin{array}{r}
\displaystyle\frac{1}{k}\left[ -\lambda _{i,n}^{L,m}\left(
p_{i,n}^{m+1}-p_{i,n}^{m}\right) +a_{i,n}^{m}\left(
q_{i,n}^{m+1}-q_{i,n}^{m}\right) \right] \hspace{1in} \\
+\frac{\lambda _{i,n}^{R,m}}{h}\left[ -\lambda _{i,n}^{L,m}\left(
p_{i,n}^{m}-p_{i,n-1}^{m}\right) +a_{i,n}^{m}\left(
q_{i,n}^{m}-q_{i,n-1}^{m}\right) \right] =d_{i,n}^{R,m}
\end{array}
\label{fde1}
\end{equation}
for $n=1,\ldots ,N$ and
\begin{equation}
\begin{array}{r}
\displaystyle\frac{1}{k}\left[ -\lambda _{i,n}^{R,m}\left(
p_{i,n}^{m+1}-p_{i,n}^{m}\right) +a_{i,n}^{m}\left(
q_{i,n}^{m+1}-q_{i,n}^{m}\right) \right] \hspace{1in} \\
+\frac{\lambda _{i,n}^{L,m}}{h}\left[ -\lambda _{i,n}^{L,m}\left(
p_{i,n+1}^{m}-p_{i,n}^{m}\right) +a_{i,n}^{m}\left(
q_{i,n+1}^{m}-q_{i,n}^{m}\right) \right] =d_{i,n}^{L,m}
\end{array}
\label{fde2}
\end{equation}
for $n=0,\ldots ,N-1$, where $a_{i,n}^{m}$, etc. are the values of the
respective functions $a_{i}$, etc. at the point $\left(
nh,mk,p_{i,n}^{m},q_{i,n}^{m}\right) $. (In this section, $n$ is always the
running index for the spatial variable, not the number of branches.) The
initial condition is simply
\begin{equation}
p_{i,n}^{0}=P_{i}^{I}\left( nh\right) ,\quad q_{i,n}^{0}=Q_{i}^{I}\left(
nh\right) .  \label{fdeic}
\end{equation}
If for a fixed $m$ the quantities $p_{i,n}^{m}$ and $q_{i,n}^{m}$ are
constructed for $n=0,\ldots ,N$, then, equations (\ref{fde1}) and (\ref{fde2}%
) determine $p_{i,n}^{m+1}$ and $q_{i,n}^{m+1}$ for $n=1,\ldots ,N-1$. The
quantities for $n=0$ and $N$ are determined by boundary conditions. At a
source end, if the boundary condition is given by (\ref{bcsp}), we impose
\begin{equation}
p_{i,0}^{m+1}=P_{i}^{B}\left( \left( m+1\right) k\right)  \label{fdebcsp}
\end{equation}
and solve $q_{i,0}^{m+1}$ from (\ref{fde2}) with $n=0$. If the boundary
condition is (\ref{bcsq}), we impose
\begin{equation}
q_{i,0}^{m+1}=Q_{i}^{B}\left( \left( m+1\right) k\right)  \label{fdebcsq}
\end{equation}
and solve $p_{i,0}^{m+1}$ from (\ref{fde2}). At a junction with $%
j_{1},\ldots ,j_{\nu }$ incoming branches and $j_{\nu +1},\ldots ,j_{\mu }$
outgoing branches, we prescribe
\begin{equation}
p_{j_{1},N}^{m+1}=p_{j_{l^{\prime }},0}^{m+1}=:p^{m+1}  \label{fdebcj1}
\end{equation}
for $l=1,\ldots ,\nu $, $l^{\prime }=\nu +1,\ldots ,\mu $, and
\begin{equation}
\sum_{l=1}^{\nu }q_{j_{l},N}^{m+1}=\sum_{l^{\prime }=\nu +1}^{\mu
}q_{j_{l^{\prime }},0}^{m+1}.  \label{fdebcj2}
\end{equation}
These equations are solved jointly with equation (\ref{fde1}) at $n=N$ for $%
i=j_{1},\ldots ,j_{\nu }$ and with equation (\ref{fde2}) at $n=0$ for $%
i=j_{\nu +1},\ldots ,j_{\mu }$. The reason that the quantities $p^{m+1}$, $%
q_{j_{l},N}^{m+1}$ and $q_{j_{l^{\prime }},0}^{m+1}$ can be uniquely solved
is that the coefficient matrix
\[
\left(
\begin{array}{ccc}
0 & R_{1} & R_{2} \\
-\frac{1}{k}S_{1} & \frac{1}{k}A_{1} & 0 \\
-\frac{1}{k}S_{2} & 0 & \frac{1}{k}A_{2}
\end{array}
\right)
\]
with
\begin{eqnarray*}
R_{1} &=&\left( 1,\ldots ,1\right) ,\quad R_{2}=\left( -1,\ldots ,-1\right) ,
\\
S_{1} &=&\left( \lambda _{j_{1},N}^{L,m},\ldots ,\lambda _{j_{\nu
},N}^{L,m}\right) ^{T},\quad S_{2}=\left( \lambda _{j_{1},0}^{R,m},\ldots
,\lambda _{j_{\nu },0}^{R,m}\right) ^{T}, \\
A_{1} &=&diag\left( a_{j_{1},N}^{m},\ldots ,a_{j_{\nu },N}^{m}\right) ,\quad
A_{2}=diag\left( a_{j_{\nu +1},0}^{m},\ldots ,a_{j_{\mu },0}^{m}\right)
\end{eqnarray*}
has the determinant
\[
\frac{1}{k^{\mu }}\left( -\sum_{l=1}^{\nu }\frac{\lambda _{j_{l},N}^{L,m}}{%
a_{j_{l},N}^{m}}+\sum_{l^{\prime }=\nu +1}^{\mu }\frac{\lambda
_{j_{l^{\prime }},0}^{R,m}}{a_{j_{l^{\prime }},0}^{m}}\right)
\prod_{l=1}^{\nu }a_{j_{l},N}^{m}\prod_{l^{\prime }=\nu +1}^{\mu
}a_{j_{l^{\prime }},0}^{m}>0.
\]
(We used the fact $\lambda _{i}^{L}<0$, $\lambda _{i}^{R}>0$ and $a_{i}>0$
here.) At a terminal end with the boundary condition (\ref{bctp}) resp. (\ref
{bctq}), we impose
\begin{equation}
p_{i,N}^{m+1}=P_{i}^{B}\left( \left( m+1\right) k\right) \quad \text{resp. }%
q_{i,N}^{m+1}=Q_{i}^{B}\left( \left( m+1\right) k\right)  \label{fdebct}
\end{equation}
and solve the other quantity from (\ref{fde1}) with $n=N$. If the boundary
condition is (\ref{bctw}), we impose
\begin{equation}
\begin{array}{r}
\displaystyle\frac{1}{k}\left( p_{i,N}^{m+1}-p_{i,N}^{m}\right) -\frac{\eta
_{i}}{k}\left( q_{i,N}^{m+1}-q_{i,N}^{m}\right) +\frac{\delta _{i}}{2}\left(
p_{i,N}^{m+1}+p_{i,N}^{m}\right) \\[12pt]
\displaystyle-\frac{\varepsilon _{i}}{2}\left(
q_{i,N}^{m+1}+q_{i,N}^{m}\right) =W_{i}^{B}\left( \left( m+\frac{1}{2}%
\right) k\right) .
\end{array}
\label{fdebctw}
\end{equation}
Together with (\ref{fde1}) for $n=N$, the values of $p_{i,N}^{m+1}$and $%
q_{i,N}^{m+1}$are uniquely determined. This is because the coefficient
matrix has the determinant
\[
\det \left(
\begin{array}{cc}
-\frac{\lambda _{i,N}^{L,m}}{k} & \frac{a_{i,N}^{m}}{k} \\
\frac{1}{k}+\frac{\delta _{i}}{2} & -\frac{\eta _{i}}{k}-\frac{\varepsilon
_{i}}{2}
\end{array}
\right) <0.
\]
(One might suspect that the simpler condition
\begin{equation}
\frac{1}{k}\left( p_{i,N}^{m+1}-p_{i,N}^{m}\right) -\frac{\eta _{i}}{k}%
\left( q_{i,N}^{m+1}-q_{i,N}^{m}\right) +\delta _{i}p_{i,N}^{m}-\varepsilon
_{i}q_{i,N}^{m}=W_{i}^{B}\left( mk\right) .  \label{fdebctw1}
\end{equation}
would also suffices. It indeed can determine unique values of $p_{i,N}^{m+1}$
and $q_{i,N}^{m+1}$. However, we are unable to prove the convergence of the
scheme with this condition. This will be clear from the proof of the next
theorem.)

It is clear that for any step-sizes $h$ and $k$, this scheme generates a
discretized solution as long as $\lambda _{i}^{L}$ remains negative at $x=0$
and $x=1$. We show that if the ratio $k/h$ is fixed and sufficiently small,
then, in a time interval the solutions for the finite-difference equations
converge to the solution to the original system of differential equations (%
\ref{depq}) as $h\rightarrow 0$.

\begin{theorem}
\label{fdeth1}Suppose that the conditions of Theorem \ref{quath1} holds and
that
\[
a_{i}\left( x,t,p,q\right) >0,\quad \lambda _{i}^{L}\left( x,t,p,q\right) <0
\]
for all $\left( x,t\right) \in \left[ 0,1\right] \times \left[ 0,\delta
\right] $ and $\left( p,q\right) \in \Bbb{R}^{2}$, where $\delta >0$ appears
in Theorem \ref{quath1}. Suppose also that the initial and boundary
functions $P_{i}^{I}$, $Q_{i}^{I}$, $P_{i}^{B}$, $Q_{i}^{B}$ and $W_{i}^{B}$
have continuous second derivatives. Let $\sigma >0$ be a positive constant
such that
\begin{equation}
\sigma \max \left\{ \left| \lambda _{i}^{L}\right| _{0,\delta },\left|
\lambda _{i}^{R}\right| _{0,\delta }\right\} <1,  \label{fdeth1.6}
\end{equation}
and let the ratio $k/h=\sigma $ be fixed. Then, there is a constant $\delta
_{0}>0$ such that, as $h\rightarrow 0$, the solutions of the
finite-difference scheme described above converges to the solution of the
differential equation (\ref{depq}) in the strip $0\leq t\leq \delta _{0}$.
\end{theorem}

\paragraph{\emph{Remark:}}

\smallskip The condition of $a_{i}>0$, $\lambda _{i}^{L}<0$ for all $\left(
p,q\right) $ is stronger than needed. One may only require that the
inequalities hold in a certain range of $\left( p,q\right) $ containing the
solution $\left( P_{i},Q_{i}\right) $ in its interior. The theorem is stated
as above to simplify the argument.

\paragraph{\noindent \noindent Proof.}

By Theorem \ref{quath1}, the system of differential equations has a solution
$\left( P_{i},Q_{i}\right) $ in $D_{\delta }$ for some $\delta >0$. Since
the initial and boundary functions have continuous second derivatives, it
can be shown using standard arguments that the solution $\left(
P_{i},Q_{i}\right) $ has continuous second order derivatives in $D_{\delta
}. $ (Reduce $\delta $ if necessary.) By Taylor's theorem and $k=\sigma h$,
we can write
\begin{equation}
\begin{array}{r}
\displaystyle\frac{1}{k}\left[ -\tilde{\lambda}_{i,n}^{L,m}\left(
P_{i,n}^{m+1}-P_{i,n}^{m}\right) +\tilde{a}_{i,n}^{m}\left(
Q_{i,n}^{m+1}-Q_{i,n}^{m}\right) \right] \hspace{1.5in} \\
+\frac{\tilde{\lambda}_{i,n}^{R,m}}{h}\left[ -\tilde{\lambda}%
_{i,n}^{L,m}\left( P_{i,n}^{m}-P_{i,n-1}^{m}\right) +\tilde{a}%
_{i,n}^{m}\left( Q_{i,n}^{m}-Q_{i,n-1}^{m}\right) \right] =\tilde{d}%
_{i,n}^{R,m}+O\left( h\right)
\end{array}
\label{fdeth1.10}
\end{equation}
for $n=1,\ldots ,N$, and
\begin{equation}
\begin{array}{r}
\displaystyle\frac{1}{k}\left[ -\tilde{\lambda}_{i,n}^{R,m}\left(
P_{i,n}^{m+1}-P_{i,n}^{m}\right) +\tilde{a}_{i,n}^{m}\left(
Q_{i,n}^{m+1}-Q_{i,n}^{m}\right) \right] \hspace{1.5in} \\
+\frac{\tilde{\lambda}_{i,n}^{L,m}}{h}\left[ -\tilde{\lambda}%
_{i,n}^{R,m}\left( P_{i,n}^{m}-P_{i,n-1}^{m}\right) +\tilde{a}%
_{i,n}^{m}\left( Q_{i,n}^{m}-Q_{i,n-1}^{m}\right) \right] =\tilde{d}%
_{i,n}^{L,m}+O\left( h\right)
\end{array}
\label{fdeth1.11}
\end{equation}
for $n=0,\ldots ,N-1$, where $P_{i,n}^{m}$ and $Q_{i,n}^{m}$ are the values
of the corresponding functions at the point $\left( nh,mk\right) $, and $%
\tilde{\lambda}_{i,n}^{L,m}$ etc. represent the values of the corresponding
functions at the point $\left( nh,mk,P_{i,n}^{m},Q_{i,n}^{m}\right) $. Let
\[
u_{i,n}^{m}=P_{i,n}^{m}-p_{i,n}^{m},\quad
v_{i,n}^{m}=Q_{i,n}^{m}-q_{i,n}^{m}.
\]
Our task is to show
\[
u_{i,n}^{m}\rightarrow 0,\quad v_{i,n}^{m}\rightarrow 0
\]
as $h\rightarrow 0$ and $k=\sigma h$. We prove it by showing that there are
positive constants $\delta _{0}$, $h_{0}$ and $M$, independent of $m$, such
that
\begin{equation}
\left| u_{i,n}^{m}\right| \leq Mh,\quad \left| v_{i,n}^{m}\right| \leq Mh,
\label{fdeth1.2}
\end{equation}
if $h\leq h_{0}$, $k=\sigma h$ and $0\leq mk\leq \delta _{0}$.

We first derive some recursive relations. Subtract (\ref{fde1}) and (\ref
{fde2}) from (\ref{fdeth1.10}) and (\ref{fdeth1.11}), respectively, and use
the Lipschitz property and the boundedness of the derivatives of $P_{i}$ and
$Q_{i}$, we obtain
\begin{equation}
\begin{array}{r}
\displaystyle\frac{1}{k}\left[ -\lambda _{i,n}^{L,m}\left(
u_{i,n}^{m+1}-u_{i,n}^{m}\right) +a_{i,n}^{m}\left(
v_{i,n}^{m+1}-v_{i,n}^{m}\right) \right] +\frac{\lambda _{i,n}^{R,m}}{h}
\left[ -\lambda _{i,n}^{L,m}\left( u_{i,n}^{m}-u_{i,n-1}^{m}\right)
+a_{i,n}^{m}\left( v_{i,n}^{m}-v_{i,n-1}^{m}\right) \right] \\
\begin{array}{ll}
= & \displaystyle O\left( h\right) +\tilde{d}_{i,n}^{R,m}-d_{i,n}^{R,m}+%
\left( \tilde{\lambda}_{i,n}^{L,m}-\lambda _{i,n}^{L,m}\right) \frac{%
P_{i,n}^{m+1}-P_{i,n}^{m}}{k}-\left( \tilde{a}_{i,n}^{m}-a_{i,n}^{m}\right)
\frac{Q_{i,n}^{m+1}-Q_{i,n}^{m}}{k} \\
& \displaystyle+\left( \tilde{\lambda}_{i,n}^{R,m}\tilde{\lambda}%
_{i,n}^{L,m}-\lambda _{i,n}^{R,m}\lambda _{i,n}^{L,m}\right) \frac{%
P_{i,n}^{m}-P_{i,n-1}^{m}}{h}-\left( \tilde{\lambda}_{i,n}^{R,m}\tilde{a}%
_{i,n}^{m}-\lambda _{i,n}^{R,m}a_{i,n}^{m}\right) \frac{%
Q_{i,n}^{m}-Q_{i,n-1}^{m}}{h} \\
= & O\left( h\right) +O\left( u_{i,n}^{m}\right) +O\left( v_{i,n}^{m}\right)
\end{array}
\end{array}
\label{fdeth1.3}
\end{equation}
and, similarly,
\begin{equation}
\begin{array}{r}
\displaystyle\frac{1}{k}\left[ -\lambda _{i,n}^{R,m}\left(
u_{i,n}^{m+1}-u_{i,n}^{m}\right) +a_{i,n}^{m}\left(
v_{i,n}^{m+1}-v_{i,n}^{m}\right) \right] +\frac{\lambda _{i,n}^{L,m}}{h}
\left[ -\lambda _{i,n}^{R,m}\left( u_{i,n+1}^{m}-u_{i,n}^{m}\right)
+a_{i,n}^{m}\left( v_{i,n+1}^{m}-v_{i,n}^{m}\right) \right] \\
=O\left( h\right) +O\left( u_{i,n}^{m}\right) +O\left( v_{i,n}^{m}\right) .
\end{array}
\label{fdeth1.4}
\end{equation}
Introduce
\[
r_{i,n}^{m}=-\lambda
_{i,n}^{L,m-1}u_{i,n}^{m}+a_{i,n}^{m-1}v_{i,n}^{m},\quad
s_{i,n}^{m}=-\lambda _{i,n}^{R,m-1}u_{i,n}^{m}+a_{i,n}^{m-1}v_{i,n}^{m}.
\]
One can show that (\ref{fdeth1.2}) is equivalent to
\begin{equation}
\left| r_{i,n}^{m}\right| \leq Mh,\quad \left| s_{i,n}^{m}\right| \leq Mh.
\label{fdeth1.1}
\end{equation}
(Throughout the proof of this theorem, we use $M$ to denote any positive
constant that is independent of $m$.) Using the identity
\[
-\lambda _{i,n}^{L,m}u_{i,l}^{m}+a_{i,n}^{m}v_{i,l}^{m}=r_{i,l}^{m}+\left(
\lambda _{i,l}^{L,m-1}-\lambda _{i,n}^{L,m}\right) u_{i,l}^{m}+\left(
a_{i,l}^{m-1}-a_{i,n}^{m}\right) v_{i,l}^{m},
\]
together with
\[
\begin{array}{r}
\lambda _{i,l}^{L,m-1}-\lambda _{i,n}^{L,m}=O\left( k\right) +O\left(
p_{i,l}^{m-1}-p_{i,n}^{m}\right) +O\left( q_{i,l}^{m-1}-q_{i,n}^{m}\right) ,
\\[12pt]
a_{i,l}^{m-1}-a_{i,n}^{m}=O\left( k\right) +O\left(
p_{i,l}^{m-1}-p_{i,n}^{m}\right) +O\left( q_{i,l}^{m-1}-q_{i,n}^{m}\right) ,
\end{array}
\]
and
\begin{eqnarray*}
p_{i,l}^{m-1}-p_{i,n}^{m} &=&-u_{i,l}^{m-1}+u_{i,n}^{m}+\left(
P_{i,l}^{m-1}-P_{i,n}^{m}\right) , \\
q_{i,l}^{m-1}-q_{i,n}^{m} &=&-v_{i,l}^{m-1}+v_{i,n}^{m}+\left(
Q_{i,l}^{m-1}-Q_{i,n}^{m}\right)
\end{eqnarray*}
for $l=n-1$, $n$, $n+1$, we can write
\[
\begin{array}{r}
-\lambda
_{i,n}^{L,m}u_{i,l}^{m}+a_{i,n}^{m}v_{i,l}^{m}=r_{i,l}^{m}+u_{i,l}^{m}O%
\left( k\right) +v_{i,l}^{m}O\left( k\right) +u_{i,l}^{m}O\left(
u_{i,l}^{m-1},u_{i,n}^{m},v_{i,l}^{m-1},v_{i,n}^{m}\right) \\
+v_{i,l}^{m}O\left(
u_{i,l}^{m-1},u_{i,n}^{m},v_{i,l}^{m-1},v_{i,n}^{m}\right) , \\
-\lambda
_{i,n}^{R,m}u_{i,l}^{m}+a_{i,n}^{m}v_{i,l}^{m}=s_{i,l}^{m}+u_{i,l}^{m}O%
\left( k\right) +v_{i,l}^{m}O\left( k\right) +u_{i,l}^{m}O\left(
u_{i,l}^{m-1},u_{i,n}^{m},v_{i,l}^{m-1},v_{i,n}^{m}\right) \\
+v_{i,l}^{m}O\left(
u_{i,l}^{m-1},u_{i,n}^{m},v_{i,l}^{m-1},v_{i,n}^{m}\right) .
\end{array}
\]
where
\[
O\left( u_{i,l}^{m-1},u_{i,n}^{m},v_{i,l}^{m-1},v_{i,n}^{m}\right) =O\left(
u_{i,l}^{m-1}\right) +O\left( u_{i,n}^{m}\right) +O\left(
v_{i,l}^{m-1}\right) +O\left( v_{i,l}^{m}\right) .
\]
Substituting these relations into (\ref{fdeth1.3}) and (\ref{fdeth1.4}), we
obtain
\begin{equation}
\begin{array}{ll}
r_{i,n}^{m+1}=r_{i,n}^{m}-\sigma \lambda _{i,n}^{R,m}\left(
r_{i,n}^{m}-r_{i,n-1}^{m}\right) +O_{i,n,n-1}^{m}, & \quad n=1,\ldots ,N, \\%
[12pt]
s_{i,n}^{m+1}=s_{i,n}^{m}-\sigma \lambda _{i,n}^{L,m}\left(
s_{i,n+1}^{m}-s_{i,n}^{m}\right) +O_{i,n,n+1}^{m}, & \quad n=0,\ldots ,N-1,
\end{array}
\label{fdeth1.9}
\end{equation}
for $m\geq 1$, where
\[
\begin{array}{r}
O_{i,n,n-1}^{m}=O\left( h^{2}\right) +h\left( O\left( u_{i,n}^{m}\right)
+O\left( v_{i,n}^{m}\right) \right) +u_{i,n-1}^{m}O\left( h\right)
+v_{i,n-1}^{m}O\left( h\right) \\[12pt]
+u_{i,n-1}^{m}O\left(
u_{i,n-1}^{m-1},u_{i,n}^{m},v_{i,n-1}^{m-1},v_{i,n}^{m}\right)
+v_{i,n-1}^{m}O\left(
u_{i,n-1}^{m-1},u_{i,n}^{m},v_{i,n-1}^{m-1},v_{i,n}^{m}\right)
\end{array}
\]
and $O_{i,n,n+1}^{m}$ is defined similarly with $n-1$ substituted by $n+1$.
These are the recursive relations we need.

We now prove (\ref{fdeth1.1}). Assume $\delta _{0}<\sigma /2$. Then, $mk\leq
\delta _{0}$ implies $m<N-m$. The proof will be divided into three cases:
(1) $m\leq n\leq N-m$, (2) $0\leq n<m$ and (3) $N-m<n\leq N$. It may be
helpful to compare the argument below with the proof of Theorem \ref{linth1}%
, in which the region $D_{i}$ is divided into $D_{i}^{C}$, $D_{i}^{L}$ and $%
D_{i}^{R}$.

\paragraph{Case 1: $m\leq n\leq N-m$.}

Let
\[
e_{m}=\max_{m\leq n\leq N-m}\left\{ \left| r_{i,n}^{m}\right| ,\left|
s_{i,n}^{m}\right| \right\} .
\]
In view of (\ref{fdeth1.6}), the coefficients of $r_{i,n}^{m}$, $%
r_{i,n-1}^{m}$, $s_{i,n}^{m}$ and $s_{i,n+1}^{m}$ in (\ref{fdeth1.9}) are
all nonnegative. Hence, from (\ref{fdeth1.9}),
\begin{equation}
e_{m+1}\leq e_{m}+C\left( h^{2}+he_{m}+e_{m}e_{m-1}+e_{m}^{2}\right) ,\quad
m\geq 1  \label{fdeth1.13}
\end{equation}
where $C>0$ is a constant. By initial condition (\ref{fdeic}),
\[
u_{i,n}^{0}=v_{i,n}^{0}=0.
\]
Thus, $e_{0}=0$. Also, by (\ref{fdeth1.9}) with $m=0$,
\begin{equation}
\begin{array}{ll}
r_{i,n}^{1}=O\left( h^{2}\right) & \quad \text{for }n=1,\ldots ,N, \\
s_{i,n}^{1}=O\left( h^{2}\right) & \quad \text{for }n=0,\ldots ,N-1.
\end{array}
\label{fdeth1.12}
\end{equation}
This implies $e_{1}=O\left( h^{2}\right) $. Consider the linear difference
equation with initial condition
\[
E_{m+1}=\left( 1+3Ch\right) E_{m}+Ch^{2},\quad m\geq 1,\quad
E_{1}=C_{0}h^{2},
\]
where $C_{0}$ is so large that $e_{1}\leq C_{0}h^{2}$. It has the solution
\begin{eqnarray*}
E_{m+1} &=&C_{0}h^{2}\left( 1+3Ch\right) ^{m}+\frac{h}{3}\left( \left(
1+3Ch\right) ^{m}-1\right) \\
&\leq &h\left( C_{0}he^{3Chm}+\frac{1}{3}e^{3Chm}-1\right) .
\end{eqnarray*}
Let $\delta _{0}$ be so small that $e^{3C\delta _{0}/\sigma }<4$. Then,
there is an $h_{0}>0$ such that $E_{m}\leq h$ for all $h\leq h_{0}$ and $%
mk\leq \delta _{0}$. This implies that
\[
E_{m+1}\geq E_{m}+C\left( h^{2}+hE_{m}+E_{m}E_{m-1}+E_{m}^{2}\right) ,\quad
E_{1}\geq e_{1}.
\]
Hence,
\[
e_{m}\leq E_{m}\leq h,
\]
which leads to (\ref{fdeth1.1}) with $M=1$ in Case 1.

\paragraph{\protect\smallskip Case 2: $0\leq n<m$.}

The proof in this case depends on the type of the boundary condition at the
left end of the branch. Suppose the end is a source with the boundary
condition (\ref{fdebcsp}). Let
\[
e_{m}=\max_{0\leq n\leq N-m}\left\{ \left| r_{i,n}^{m}\right| ,\left|
s_{i,n}^{m}\right| \right\} .
\]
(As was the case in the proof of Theorem \ref{linth1}, it is more convenient
to include the central trapezoidal part $m\leq n\leq N-m$.) Hence, from (\ref
{fdeth1.9})
\begin{equation}
\begin{array}{r}
\left| r_{i,n}^{m+1}\right| \leq \left| e_{m}\right| +C\left(
h^{2}+he_{m}+e_{m}e_{m-1}+e_{m}^{2}\right) \quad \text{for }n=1,\ldots ,N-m,
\\[12pt]
\left| s_{i,n}^{m+1}\right| \leq \left| e_{m}\right| +C\left(
h^{2}+he_{m}+e_{m}e_{m-1}+e_{m}^{2}\right) \quad \text{for }n=0,\ldots ,N-m.
\end{array}
\label{fdeth1.14}
\end{equation}
Since by (\ref{fdebcsp}), $u_{i,0}^{m}=0$, it follows that $%
r_{i,0}^{m}=s_{i,0}^{m}$ for all $m$. Therefore, $e_{m}$ satisfies the same
difference inequality (\ref{fdeth1.13}). We also have $e_{1}=O\left(
h^{2}\right) $ by (\ref{fdeth1.12}). Thus, the above analysis gives $%
e_{m}\leq h$.

Suppose the boundary condition is given by (\ref{fdebcsq}), then, $%
v_{i,0}^{m}=0$ and
\[
r_{i,0}^{m}=\frac{\lambda _{i,0}^{L,m-1}}{\lambda _{i,0}^{R,m-1}}s_{i,0}^{m}
\]
for all $m\geq 1$. Let $\hat{r}_{i,n}^{m}=r_{i,n}^{m}/M$ where $M$ is
sufficiently large such that
\[
M>\max_{m}\left\{ \left| \frac{\lambda _{i,0}^{L,m}}{\lambda _{i,0}^{R,m}}%
\right| \right\} .
\]
Then, (\ref{fdeth1.9}) still holds with $r$ substituted by $\hat{r}$. Let
\[
e_{m}=\max_{0\leq n\leq N-m}\left\{ \left| \hat{r}_{i,n}^{m}\right| ,\left|
s_{i,n}^{m}\right| \right\} .
\]
We again have (\ref{fdeth1.14}) and
\[
\left| \hat{r}_{i,0}^{m+1}\right| \leq \left| s_{i,0}^{m+1}\right| \leq
\left| e_{m}\right| +C\left( h^{2}+he_{m}+e_{m}e_{m-1}+e_{m}^{2}\right) .
\]
Hence, $e_{m}$ satisfies (\ref{fdeth1.13}) again. Therefore,
\[
\left| r_{i,n}^{m}\right| \leq Mh,\quad \left| s_{i,n}^{m}\right| \leq h.
\]

Suppose the left end is a junction. We shall treat all the branches
connected to the same junction simultaneously. Let $j_{1},\ldots ,j_{\nu }$
be the incoming branches and $j_{\nu +1},\ldots ,j_{\mu }$ the outgoing
branches. It is easy to see that the boundary conditions (\ref{fdebcj1})--(%
\ref{fdebcj2}) are satisfied if $p$ and $q$ are substituted by $u$ and $v$,
respectively. Using the identities
\begin{equation}
u_{i,n}^{m+1}=\frac{r_{i,n}^{m+1}-s_{i,n}^{m+1}}{\lambda _{i,n}^{m}},\quad
v_{i,n}^{m+1}=\frac{\lambda _{i,n}^{R,m}r_{i,n}^{m+1}-\lambda
_{i,n}^{L,m}s_{i,n}^{m+1}}{a_{i,n}^{m}\lambda _{i,n}^{m}},  \label{fdeth1.16}
\end{equation}
where
\[
\lambda _{i,n}^{m}=\lambda _{i,n}^{R,m}-\lambda _{i,n}^{L,m}>0,
\]
the equations for $r$ and $s$ have the form
\[
\begin{array}{l}
\begin{array}{ll}
\frac{1}{\lambda _{j_{1},N}^{m}}\left(
r_{j_{1},N}^{m+1}-s_{j_{1},N}^{m+1}\right) -\frac{1}{\lambda _{i,N}^{m}}%
\left( r_{i,N}^{m+1}-s_{i,N}^{m+1}\right) =0, & \quad i=j_{2},\ldots ,j_{\nu
}, \\
\frac{1}{\lambda _{j_{1},N}^{m}}\left(
r_{j_{1},N}^{m+1}-s_{j_{1},N}^{m+1}\right) -\frac{1}{\lambda _{i,0}^{m}}%
\left( r_{i,0}^{m+1}-s_{i,0}^{m+1}\right) =0, & \quad i=j_{\nu +1},\ldots
,j_{\mu },
\end{array}
\\
\sum_{l=1}^{\nu }\frac{1}{a_{j_{l},N}^{m}\lambda _{j_{l},N}^{m}}\left(
\lambda _{j_{l},N}^{R,m}r_{j_{l},N}^{m+1}-\lambda
_{j_{l},N}^{L,m}s_{j_{l},N}^{m+1}\right) -\sum_{l^{\prime }=\nu +1}^{\mu }%
\frac{1}{a_{j_{l^{\prime }},0}^{m}\lambda _{j_{l^{\prime }},0}^{m}}\left(
\lambda _{j_{l^{\prime }},0}^{R,m}r_{j_{l^{\prime }},0}^{m+1}-\lambda
_{j_{l^{\prime }},0}^{L,m}s_{j_{l^{\prime }},0}^{m+1}\right) =0.
\end{array}
\]
The system can be solved for $s_{j_{1},N}^{m+1},\ldots ,s_{j_{\nu },N}^{m+1}$%
, $r_{j_{\nu +1},0}^{m+1},\ldots ,r_{j_{\mu },0}^{m+1}$ because the
coefficient matrix
\[
\left(
\begin{array}{cccc}
-\frac{1}{\lambda _{j_{1},N}^{m}} & \frac{1}{\lambda _{j_{2},N}^{m}} & \cdots
& 0 \\
\vdots & \vdots & \ddots & \vdots \\
-\frac{1}{\lambda _{j_{1},N}^{m}} & 0 & \cdots & -\frac{1}{\lambda _{j_{\mu
},0}^{m}} \\
-\frac{\lambda _{j_{1,N}}^{L,m}}{\lambda _{j_{1},N}^{m}a_{j_{1},N}^{m}} & -%
\frac{\lambda _{j_{2},N}^{L,m}}{\lambda _{j_{2},N}^{m}a_{j_{2},N}^{m}} &
\cdots & -\frac{\lambda _{j_{\mu },0}^{R,m}}{\lambda _{j_{\mu
},0}^{m}a_{j_{\mu },0}^{m}}
\end{array}
\right)
\]
has the determinant
\[
\frac{\left( -1\right) ^{\nu +1}}{\prod_{l=1}^{\nu }\lambda
_{j_{l},N}^{m}\prod_{l^{\prime }=\nu +1}^{\mu }\lambda _{j_{l^{\prime
}},0}^{m}}\left( -\sum_{l=1}^{\nu }\frac{\lambda _{j_{l},N}^{L,m}}{%
a_{j_{l},N}^{m}}+\sum_{l^{\prime }=\nu +1}^{\mu }\frac{\lambda
_{j_{l^{\prime }},0}^{R,m}}{a_{j_{l^{\prime }},N}^{m}}\right) \neq 0.
\]
(We used here $\lambda _{i,n}^{m}>0$, $a_{i,n}^{m}>0$, $\lambda
_{i,n}^{R,m}>0$ and $\lambda _{i,n}^{L,m}<0$.) Let the solution be written
as
\begin{equation}
\begin{array}{ll}
\displaystyle s_{i,N}^{m+1}=\sum_{l=1}^{\nu
}m_{j_{l}}^{i}r_{j_{i},N}^{m+1}+\sum_{l^{\prime }=\nu +1}^{\mu
}m_{j_{l^{\prime }}}^{i}s_{j_{l^{\prime }},0}^{m+1}, & \quad i=j_{1},\ldots
,j_{\nu }, \\
\displaystyle r_{i,0}^{m+1}=\sum_{l=1}^{\nu
}n_{j_{l}}^{i}r_{j_{i},N}^{m+1}+\sum_{l^{\prime }=\nu +1}^{\mu
}n_{j_{l^{\prime }}}^{i}s_{j_{l^{\prime }},0}^{m+1}, & \quad i=j_{\nu
+1},\ldots ,j_{\mu }.
\end{array}
\label{fdeth1.15}
\end{equation}
Choose a constant $M$ such that
\[
M>\max_{i=j_{1},\ldots j_{\mu }}\left\{ \sum_{l=1}^{\mu }\left|
m_{j_{l}}^{i}\right| ,\sum_{l=1}^{\mu }\left| n_{j_{l}}^{i}\right| \right\}
\]
and introduce
\[
\hat{s}_{j_{l},n}^{m}=s_{j_{l},n}^{m}/M,\quad \hat{r}_{j_{l^{\prime
}},n}^{m}=r_{j_{l^{\prime }},n}^{m}/M
\]
for $l=1,\ldots ,\nu $, $l^{\prime }=\nu +1,\ldots ,\mu $. Equations in (\ref
{fdeth1.9}) still hold if $\hat{s}_{j_{l},n}^{m}$ and $\hat{r}_{j_{l^{\prime
}},n}^{m}$ are substituted for $s_{j_{l},n}^{m}$ and $r_{j_{l^{\prime
}},n}^{m}$, respectively. Let $e_{m}$ denote the maximum of the quantities
\[
\max\Sb m\leq n\leq N  \\ 1\leq l\leq \nu  \endSb \left\{ \left|
r_{j_{l},n}^{m}\right| ,\left| \hat{s}_{j_{l},n}^{m}\right| \right\} ,\quad
\max\Sb 0\leq n\leq N-m  \\ \nu +1\leq l^{\prime }\leq \mu  \endSb \left\{
\left| \hat{r}_{j_{l^{\prime }},n}^{m}\right| ,\left| s_{j_{l^{\prime
}},n}^{m}\right| \right\} .
\]
(Notice again the inclusion of the middle part $m\leq n\leq N-m$.) Since the
coefficients of $r$ and $s$ are all positive, it is easy to see that
\[
\left| r_{j_{l},n}^{m+1}\right| \leq e_{m}+C\left(
h^{2}+he_{m}+e_{m}e_{m-1}+e_{m}^{2}\right)
\]
for $l=1,\ldots ,\nu $, $n=m,\ldots ,N$ and
\[
\left| s_{j_{l^{\prime }},n}^{m+1}\right| \leq e_{m}+C\left(
h^{2}+he_{m}+e_{m}e_{m-1}+e_{m}^{2}\right)
\]
for $l=\nu +1,\ldots ,\mu $, $n=0,\ldots ,m$. Similar inequalities can be
derived for $\left| \hat{s}_{j_{l},n}^{m+1}\right| $, $l=1,\ldots ,\nu $, $%
n=m,\ldots ,N-1$ and for $\left| \hat{r}_{j_{l^{\prime }},n}^{m+1}\right| $,
$l^{\prime }=\nu +1,\ldots ,\mu $, $n=1,\ldots ,m$. Furthermore, by (\ref
{fdeth1.15})
\begin{eqnarray*}
\left| \hat{s}_{j_{l},N}^{m+1}\right| &=&\frac{1}{M}\left| \sum_{l=1}^{\nu
}m_{j_{l}}^{i}r_{j_{i},N}^{m+1}+\sum_{l^{\prime }=\nu +1}^{\mu
}m_{j_{l^{\prime }}}^{i}s_{j_{l^{\prime }},0}^{m+1}\right| \leq \max\Sb %
1\leq l\leq \nu  \\ \nu +1\leq l^{\prime }\leq \mu  \endSb \left\{ \left|
r_{j_{l},N}^{m+1}\right| ,\left| s_{j_{l^{\prime }},0}^{m+1}\right| \right\}
, \\
\left| \hat{r}_{j_{l},N}^{m+1}\right| &=&\frac{1}{M}\left| \sum_{l=1}^{\nu
}n_{j_{l}}^{i}r_{j_{i},N}^{m+1}+\sum_{l^{\prime }=\nu +1}^{\mu
}n_{j_{l^{\prime }}}^{i}s_{j_{l^{\prime }},0}^{m+1}\right| \leq \max\Sb %
1\leq l\leq \nu  \\ \nu +1\leq l^{\prime }\leq \mu  \endSb \left\{ \left|
r_{j_{l},N}^{m+1}\right| ,\left| s_{j_{l^{\prime }},0}^{m+1}\right| \right\}
.
\end{eqnarray*}
Therefore, we achieve again the difference inequality (\ref{fdeth1.13}) for $%
e_{m}$. Hence, $e_{m}\leq h$, and consequently,
\[
\left| r_{i,n}^{m}\right| \leq h,\quad \left| s_{i,n}^{m}\right| \leq Mh.
\]
This not only proves (\ref{fdeth1.1}) for Case 2, but also for the part of
Case 3 where the right endpoint is a junction.

\paragraph{Case 3: $N-m\leq n\leq N$.}

It only remains to discuss the case where the right end is a terminal. If
the boundary condition is given by (\ref{fdebct}), the results follow from
similar arguments in Case 2, when the source end boundary condition is
either (\ref{fdebcsp}) or (\ref{fdebcsq}). Thus, we shall only discuss the
case when the boundary condition is given by (\ref{fdebctw}), which
corresponds to the windkessel-type boundary condition (\ref{bctw}) for the
differential equations.

From (\ref{bctw}), we derive
\[
\begin{array}{r}
\displaystyle\frac{1}{k}\left( P_{i,N}^{m+1}-P_{i,N}^{m}\right) -\frac{\eta
_{i}}{k}\left( Q_{i,N}^{m+1}-Q_{i,N}^{m}\right) +\frac{\delta _{i}}{2}\left(
P_{i,N}^{m+1}+P_{i,N}^{m}\right) \\[12pt]
\displaystyle-\frac{\varepsilon _{i}}{2}\left(
Q_{i,N}^{m+1}+Q_{i,N}^{m}\right) =W_{i}^{B}\left( \left( m+\frac{1}{2}%
\right) k\right) +O\left( k^{2}\right) .
\end{array}
\]
Subtracting (\ref{fdebctw}) from above yields
\[
\frac{1}{k}\left( u_{i,N}^{m+1}-u_{i,N}^{m}\right) -\frac{\eta _{i}}{k}%
\left( v_{i,N}^{m+1}-v_{i,N}^{m}\right) +\frac{\delta _{i}}{2}\left(
u_{i,N}^{m+1}+u_{i,N}^{m}\right) -\frac{\varepsilon _{i}}{2}\left(
v_{i,N}^{m+1}+v_{i,N}^{m}\right) =O\left( k^{2}\right) .
\]
Let
\[
f^{m}=\left( 1+\frac{\delta _{i}k}{2}\right) u_{i,N}^{m}-\left( \eta _{i}+%
\frac{\varepsilon _{i}k}{2}\right) v_{i,N}^{m},\quad m=0,1,\ldots .
\]
The equation for $f^{m}$ has the form
\[
f^{m+1}=f^{m}+k\left( \varepsilon _{i}v_{i,N}^{m}-\delta
_{i}u_{i,N}^{m}\right) +O\left( k^{3}\right) .
\]
Since $f^{0}=0$, the difference equation has the solution
\[
f^{m+1}=k\sum_{j=0}^{m}\left( \varepsilon _{i}v_{i,N}^{j}-\delta
_{i}u_{i,N}^{j}\right) +O\left( k^{2}\right) .
\]
From (\ref{fdeth1.16}), we obtain
\begin{equation}
s_{i,N}^{m+1}=\frac{M_{i}^{m}}{N_{i}^{m}}r_{i,N}^{m+1}-\frac{k}{N_{i}^{m}}%
\sum_{j=0}^{m}\left( \varepsilon _{i}v_{i,N}^{j}-\delta
_{i}u_{i,N}^{j}\right) +O\left( k^{2}\right)  \label{fdeth1.17}
\end{equation}
where
\begin{eqnarray*}
M_{i}^{m} &=&\frac{1}{\lambda _{i,n}^{m}}\left( 1+\frac{\delta _{i}k}{2}%
-\left( \eta _{i}+\frac{\varepsilon _{i}k}{2}\right) \frac{\lambda
_{i,n}^{R,m}}{a_{i,n}^{m}}\right) , \\
N_{i}^{m} &=&\frac{1}{\lambda _{i,n}^{m}}\left( 1+\frac{\delta _{i}k}{2}%
-\left( \eta _{i}+\frac{\varepsilon _{i}k}{2}\right) \frac{\lambda
_{i,n}^{L,m}}{a_{i,n}^{m}}\right) .
\end{eqnarray*}
(Notice that $N_{i}^{m}>0$, hence (\ref{fdeth1.17}) is valid.) Let $\hat{s}%
_{i,n}^{m}=s_{i,n}^{m}/M$ where $M$ is a constant to be determined later.
Also let
\[
e_{m}=\max\Sb m\leq n\leq N  \\ 0\leq j\leq m  \endSb \left\{ \left|
r_{i,n}^{j}\right| ,\left| \hat{s}_{i,n}^{j}\right| \right\} .
\]
Unlike previous cases where $e_{m}$ depends on the $m$-th level quantities,
here it is more convenient to let $e_{m}$ be the maximum of all the lower
level quantities. Then, by (\ref{fdeth1.9}) modified with $\hat{s}$
substituted for $s$,
\begin{equation}
\left| r_{i,n}^{m+1}\right| \leq e_{m}+C\left(
h^{2}+he_{m}+e_{m}e_{m-1}+e_{m}^{2}\right)  \label{fdeth1.18}
\end{equation}
for $n=m,\ldots ,N$ and
\[
\left| \hat{s}_{i,n}^{m+1}\right| \leq e_{m}+C\left(
h^{2}+he_{m}+e_{m}e_{m-1}+e_{m}^{2}\right)
\]
for $n=m,\ldots ,N-1$, where $C$ is a positive constant. Also, by (\ref
{fdeth1.17}) and the relation $mk\leq \delta _{0}$,
\[
\left| \hat{s}_{i,N}^{m+1}\right| \leq \frac{1}{M}\left| \frac{M_{i}^{m}}{%
N_{i}^{m}}\right| \left| r_{i,N}^{m+1}\right| +\delta _{0}C^{\prime
}e_{m}+O\left( h^{2}\right)
\]
where $C^{\prime }>0$ is constant. Hence, from (\ref{fdeth1.18}) we see that
if $M$ is sufficiently large and $\delta _{0}$ is sufficiently small, we can
ensure
\[
\left| \hat{s}_{i,N}^{m+1}\right| \leq e_{m}+C\left(
h^{2}+he_{m}+e_{m}e_{m-1}+e_{m}^{2}\right) .
\]
(This is where the boundary condition (\ref{fdebctw1}) fails. Instead of $%
O\left( h^{2}\right) $, it can only provide $O\left( h\right) $, which is
inconsistent with (\ref{fdeth1.13}).) Thus, $e_{m}$ satisfies the relation (%
\ref{fdeth1.13}), which leads to $e_{m}\leq h$. We have thus shown that
\[
\left| r_{i,n}^{m}\right| \leq h,\quad \left| s_{i,n}^{m}\right| \leq Mh.
\]
This completes the proof of Case 3, and also the entire theorem. \endproof

\section{Discussion\label{conc}}

We have given a rather thorough treatment to the initial-boundary value
problem of the first-order quasilinear system (\ref{depq}) with various
source and terminal boundary conditions. From our results, it can be seen
that the junction condition (\ref{bcj}), which stems from the conservation
of mass and Navier-Stokes momentum, is consistent with the differential
equations. Also, the windkessel-type terminal boundary condition does not
cause problems to the solvability. However, due to the nature of the
first-order hyperbolic equations, the existence of global solution generally
is not guaranteed. This problem may disappear if more accurate models are
used. For example, in (\ref{depq}) and its special case (\ref{deaq}), only
the effect of viscosity on the wall of the vessels is taken into
consideration. If we include viscosity more comprehensively, a term of $\mu
\nabla ^{2}Q_{i}$ appears in the right side of the second equations of (\ref
{depq}) and (\ref{deaq}). The system then becomes parabolic, instead of
hyperbolic. It is well-known that parabolic systems have better regularity
properties than hyperbolic ones. Therefore, it may be possible to prove the
existence of global solutions. We are currently investigating this issue.

We have developed a numerical scheme for the computation of solutions and
proved its convergence. Although our scheme uses a nonstaggered method
similar to the one developed by Raines, \emph{et al} \cite{RJS71,RJS74},
they are substantially different. (By nonstaggered, we mean the values of $%
P_{i}$ and $Q_{i}$ are approximated at the same mesh points, unlike the
staggered method developed in \cite{CK78,KC85}.) This is because ours is
based on the normal form of the equations and takes into account of the
characteristic directions. This may explain why our scheme converges even if
the network has loops while the other can break down (cf. \cite{KC85}).

\paragraph{\noindent Acknowledgment.}

\textbf{\ }Weihua Ruan is partially supported by VasSol, Inc.

\end{document}